%% file: main.tex
\documentclass[11pt]{article}

\usepackage{empheq}
\usepackage{amsmath,amsfonts,amssymb,amsbsy,amscd}
\usepackage{mathrsfs,array}
\usepackage{subcaption}
\usepackage[a4paper, total={6in, 8in}]{geometry}
\usepackage{amsthm}

\newtheorem{theorem}{Theorem}[section]
\newtheorem{lemma}[theorem]{Lemma}
\newtheorem{proposition}[theorem]{Proposition}%
\newtheorem{remark}[theorem]{Remark}%

\usepackage{tikz}
\usetikzlibrary{arrows.meta}
\usetikzlibrary{3d}
\usetikzlibrary{shadings}
\usetikzlibrary{math}
\usepackage{adjustbox}\usepackage{multirow}
\usepackage{upgreek}
\usetikzlibrary{intersections, arrows, automata, backgrounds, calendar, chains, matrix, mindmap, patterns, petri, shadows, shapes.geometric, shapes.misc, spy, trees, shapes}

\definecolor{darkgreen}{rgb}{0, 0.7, 0}
\definecolor{orange}{rgb}{0.98, 0.6, 0.01}
 	\definecolor{napiergreen}{rgb}{0.16, 0.5, 0.0}

\usepackage{upgreek}

\newcolumntype{R}[2]{%
    >{\adjustbox{angle=#1,lap=\width-(#2)}\bgroup}%
    l%
    <{\egroup}%
}
\makeatletter
\newcommand{\thickhline}{%
    \noalign {\ifnum 0=`}\fi \hrule height 1pt
    \futurelet \reserved@a \@xhline
}
\newcolumntype{"}{@{\hskip\tabcolsep\vrule width 1pt\hskip\tabcolsep}}
\makeatother

\usepackage{pifont}
\newcommand{\cmark}{\ding{51}}%
\newcommand{\xmark}{\ding{55}}%

\usepackage{hyperref}
\hypersetup{
    colorlinks=true,
    linkcolor=blue,
    urlcolor=red,
    linktoc=all,
    allcolors=blue
}

\usepackage[nameinlink,capitalise,noabbrev]{cleveref}
\makeatletter
\newcommand{\myref}[1]{\cref{#1}\mynameref{#1}{\csname r@#1\endcsname}}
\newcommand{\Myref}[1]{\Cref{#1}\mynameref{#1}{\csname r@#1\endcsname}}

\usepackage{accents}
\newlength{\dhatheight}

\allowdisplaybreaks

\title{Thermodynamically consistent diffuse-interface mixture models of incompressible multicomponent fluids}
\author{M.F.P. ten Eikelder$^{\dag,}$\thanks{Corresponding author. e-mail: \texttt{marco.eikelder@tu-darmstadt.de}}
\and K.G. van der Zee$^\ddag$
\and D. Schillinger$^\dag$
}
\date{%
    $^\dag$Institute for Mechanics, Computational Mechanics Group, Technical University of Darmstadt\\
    $^\ddag$School of Mathematical Sciences, University of Nottingham
}

\input{declarations}

\begin{document}

\maketitle

\begin{abstract}
In this paper we derive a class of thermodynamically consistent diffuse-interface mixture models of incompressible multicomponent fluids. The class of mixture models is fully compatible with the continuum theory of mixtures. The resulting mixture models may be formulated either in constituent or in mixture quantities. This permits a direct comparison with the Navier-Stokes Cahn-Hilliard model with non-matching densities, which reveals the key modeling simplifications of the latter.
\end{abstract}




\section{Introduction}
\subsection{Background}
The description of diffuse-interface multi-constituent flows in which the interface has a positive thickness may be traced back to Rayleigh \cite{rayleigh1892xix} and van der Waals \cite{waals1894}. Based on these works, the pioneering work of Korteweg \cite{korteweg1901forme} and others, diffuse-interface models governing the motion of multiple constituents (fluids) or phases have been developed \cite{anderson1998diffuse,oden2010general} and applied in computations \cite{yue2004diffuse,gomez2018computational,ten2021novel}. In the scenario of multi-phase flow, the prototypical model is the Navier-Stokes-Korteweg model. On the other hand, mixture theory of rational mechanics provides the theoretical framework of the dynamics of multi-constituent mixtures. The first contributions on simple mixtures are the works of Fick \cite{fick1855} and Darcy \cite{darcy1856fontaines}. Since then, the topic has become more mature with the important contributions of Truesdell \cite{truesdell1957,truesdell1962mechanical} and Truesdell and Toupin \cite{truesdell1960classical}. More complete overviews of rational mixture theory are provided by Green and Naghdi \cite{green1967theory}, M\"{u}ller \cite{muller1975thermo}, M\"{u}ller and Ruggeri \cite{muller2013rational}, Bowen \cite{bowen1980incompressible,bowen1982compressible}, Truesdell \cite{truesdell1984historical}, Morro \cite{morro2016nonlinear}, and others. 

The study of incompressible diffuse-interface multi-fluid models seems only weakly connected with continuum mixture theory. Indeed, the study of diffuse-interface multi-fluid models was initiated in 1970 independent of the continuum theory of mixtures. In that year Hohenberg and Halperin proposed a model, known as \textit{model H}, for the coupling of viscous fluid incompressible flow and spinoidal decomposition \cite{hohenberg1977theory}. This diffuse-interface model is now recognized as the first \textit{Navier-Stokes Cahn-Hilliard} (NSCH) model. As the name suggests, the model is presented as the coupling between the incompressible (isothermal) Navier-Stokes equations and (an extension of) the Cahn-Hilliard equation. The capillary forces are modeled through the introduction of an additional Korteweg-type contribution to the stress tensor. Model H was initially established via phenomenological arguments, and a continuum mechanics derivation was presented by Gurtin \cite{gurtinmodel}. This derivation, and the resulting model are not compatible with the continuum theory of mixtures.

The major assumption in model H is the constant density of the mixture as well as of the individual constituents (making it not applicable to problems with large density ratios). This limitation initiated the generalization of model H to NSCH models with non-matching densities. Noteworthy contributions include the models of Lowengrub and Truskinovsky \cite{lowengrub1998quasi}, Boyer \cite{boyer2002theoretical}, Ding et al.  \cite{ding2007diffuse}, Abels et al. \cite{abels2012thermodynamically}, Shen et al. \cite{shen2013mass}, Aki et al. \cite{aki2014quasi} and Shokrpour Roudbari et al. \cite{shokrpour2018diffuse}. These models all aim to describe the same physical phenomena (the evolution of isothermal incompressible mixtures), yet they are (seemingly) distinct from one another.

In a recent article we have proposed a unified framework of all existing Navier-Stokes Cahn-Hilliard models with non-matching densities and non-zero mass fluxes \cite{eikelder2023unified}. In this work we have established one NSCH system of balance laws and have shown that many alternate forms of the same model are connected via variable transformations. As such, in this paper we no longer think of a wide variety of NSCH models, but instead of \textit{the NSCH model} (variations only occur in constitutive modeling). A particular formulation of the NSCH model reads:
\begin{subequations}\label{eq: NSCH  model intro}
  \begin{align}
  \partial_t (\rho \bv) + {\rm div} \left( \rho \bv\otimes \bv \right) + \nabla p + {\rm div} \left( \nabla \phi \otimes \dfrac{\partial  \bar{\Psi}}{\partial \nabla \phi} + (\bar{\mu}\phi-\bar{\Psi})\mathbf{I} \right) & \nn\\
    - {\rm div} \left(   \nu (2\mathbf{D}+\lambda({\rm div}\bv) \mathbf{I}) \right)-\rho\mathbf{b} &=~ 0, \label{eq: NSCH model v intro}\\
 \partial_t \rho + {\rm div}(\rho \bv) &=~ 0, \label{eq: NSCH model rho intro}\\
   \partial_t \phi + {\rm div}(\phi \bv) - {\rm div} \left(\bar{\mathbf{M}}\nabla (\bar{\mu}+\omega p)\right) +\zeta \bar{m} (\bar{\mu} + \omega p) &=~0,\label{eq: NSCH model phi intro}\\
  \bar{\mu} - \dfrac{\partial \bar{\Psi}}{\partial \phi}+{\rm div} \left(  \dfrac{\partial \bar{\Psi}}{\partial \nabla \phi} \right)&=~0.\label{eq: NSCH model mu intro}
  \end{align}
\end{subequations}
Here $\rho$ is the mixture density, $\bv$ the mixture velocity, $p$ the pressure, $\phi$ an order parameter and $\bar{\mu}$ a chemical potential quantity. Furthermore, $\bar{\mathbf{M}}=\bar{\mathbf{M}}(\phi,\nabla \phi, \bar{\mu}, \nabla \bar{\mu}, p)$ and $\bar{m}=\bar{m}(\phi, \bar{\mu}, p)$ are degenerate mobilities, $\nu$ the dynamic viscosity of the mixture, $\bg$ the gravitational acceleration, $\rho_1$ and $\rho_2$ constant specific densities of the constituents, $\omega = (\rho_2-\rho_1)/(\rho_1+\rho_2)$, and $\zeta = (\rho_1+\rho_2)/(2\rho_1\rho_2)$. We provide precise definitions in \cref{sec: connection}.

\subsection{Objective and main results}
The unified framework presented in ten Eikelder et al. \cite{eikelder2023unified} completes the fundamental exploration of alternate non-matching density NSCH models. However, the NSCH model is not compatible with mixture theory of rational mechanics. Namely, in the construction of the NSCH model, the evolution equation of the diffusive flux that results from mixture theory is replaced by a constitutive model. Therefore, the NSCH model may be classified as a \textit{reduced mixture model}.
This observation bring us to the main objective of this article: \textit{to derive a thermodynamically-consistent diffuse-interface incompressible mixture model compatible with continuum mixture theory}. We restrict to isothermal constituents. The thermodynamically-consistent property of the mixture model refers to the compatibility with the second law of thermodynamics. In particular, we derive the following mixture model:
\begin{subequations}\label{eq: model FE intro}
  \begin{align}
 \partial_t \trho_\mA + {\rm div}(\trho_\mA \bv_\mA) -\hat{\gamma}_\mA &=~ 0, \label{eq: model FE: cont intro}\\
 \partial_t (\trho_\mA \bv_\mA) + {\rm div} \left( \trho_\mA \bv_\mA\otimes \bv_\mA \right)  + \phi_\mA\nabla \left(p + \mu_\mA \right)  \nn\\
 - {\rm div}\left(\tilde{\nu}_\mA \left(2\mathbf{D}_\mA + \lambda_\mA {\rm div}\mathbf{v}_\mA\right)\right) -\trho_\mA\mathbf{b}&\nn\\
    -\displaystyle\sum_\mB \dfrac{p\phi_\mA\phi_\mB}{D_{\mA\mB}}(\bv_\mB-\bv_\mA)- \boldsymbol{\beta}_\mA&=~ 0, \label{eq: model FE: mom intro}
  \end{align}
\end{subequations}
for $\mA = 1,...,N$. Here $\trho_\mA$ is the partial mass density of constituent $\mA$, $\bv_\mA$ the constituent velocity, $\phi_\mA$ the constituent volume fraction, and $\mu_\mA$ a constituent chemical potential. Furthermore, the model contains two distinct pressure quantities, $\pi_\mA$ is the thermodynamical pressure of constituent $\mA$ and $p$ the mechanical pressure of the mixture. Finally, $\nu_\mA$ is the constituent dynamical viscosity, $\mathbf{D}_\mA$ the constituent symmetric velocity gradient, $D_{\mA \mB}$ a diffusion coefficient associated with constituents $\mA$ and $\mB$, and $\hat{\gamma}_\mA$ and $\boldsymbol{\beta}_\mA$ mass transfer (related) terms. We provide precise definitions in \cref{sec: 2nd law,sec: diffuse interface}.

The distinguishing feature of the model lies in the occurrence of both a mass and a momentum balance equation per constituent. Reduced models (e.g. NSCH and Navier-Stokes Allen-Cahn) typically contain a phase equation per constituent but a single momentum equation for the mixture. This decrease in complexity comes at the cost of violating mixture theory of rational mechanics. Another interesting aspect is that the model has no Cahn-Hilliard type equation. Furthermore we note the presence of the multiple pressure quantities. The single mechanical pressure variable $p$ acts as a Lagrange multiplier of the mixture incompressibility constraint. On the other hand, the thermodynamical pressure $\pi_\mA$ is solely associated with constituent $\mA$. 
The last line in the constituent momentum equations models the momentum transfer between the constituents. As such, we observe that constituent momentum interaction is absent in the Stefan-Maxwell equilibrium balance. Another important feature of the model is that  the equilibrium profile coincides with that of the NSCH model (for the standard Ginzburg-Landau free energy).

\subsection{Plan of the paper}
The remainder of the paper is structured as follows. In \cref{sec: mix theory} we present the general continuum theory of incompressible fluid mixtures. Here we present identities that relate constituent and mixture quantities. We exclude thermal effects. Next, in \cref{sec: 2nd law} we perform constitutive modeling via the Coleman-Noll procedure. Then, in \cref{sec: diffuse interface} we present particular diffuse-interface models. We compare the resulting models with the NSCH model in \cref{sec: connection}. Finally, in \cref{sec: discussion} we conclude and outline avenues for future research.



\section{Continuum theory of mixtures}\label{sec: mix theory}

The purpose of this section is to lay down the continuum theory of mixtures composed of incompressible isothermal constituents. The theory is based on three metaphysical principles proposed in the groundbreaking works of Truesdell and Toupin \cite{truesdell1960classical}:
\begin{enumerate}
    \item \textit{All properties of the mixture must be mathematical consequences of properties of the constituents.}
\item \textit{So as to describe the motion of a constituent, we may in imagination isolate it from the rest of the mixture, provided we allow properly for the actions of the other constituents upon it.}
\item \textit{The motion of the mixture is governed by the same equations as is a single body.}
\end{enumerate}
The first principle states that the mixture is composed of its constituent parts. The second principle asserts the physics model to be band together via interaction flux, forces or energies. Finally, the third principle ensures that the motion of a mixture is indistinguishable from that of a single fluid.

In \cref{sec: prelim} we introduce the fundamentals of the continuum theory of mixtures and the necessary kinematics. Then, in \cref{sec: BL} we provide balance laws of individual constituents and associated mixtures.

\subsection{Preliminaries and kinematics}\label{sec: prelim}

The core idea of the continuum theory of mixtures is that the material body $\mathscr{B}$ is composed of $N$ constituent bodies $\mathscr{B}_\mA$, with $\mA = 1, \dots, N$. The bodies $\mathscr{B}_\mA$ are allowed to occupy, simultaneously, a common region in space. Denote with $\mathbf{X}_{\mA}$ the spatial position of a particle of $\mathscr{B}_\mA$ in the Lagrangian (reference) configuration. The spatial position of a particle is given by the (invertible) deformation map
\begin{align}
    \mathbf{x} := \bchi_{\mA}(\mathbf{X}_{\mA},t). 
\end{align}
Consider from now on positions $\mathbf{x}$ that are taken by one particle from each of the $N$ constituent bodies $\mathscr{B}_\mA$. Around this spatial position $\mathbf{x}$ we consider an arbitrary mixture control volume $V \subset \Omega$ with measure $\vert V \vert$. Furthermore, we introduce volume $V_{\mA} \subset V$, with measure $\vert V_{\mA}\vert$, as the control volume of constituent $\mA$. The constituents  masses denote $M_{\mA}=M_{\mA}(V)$ and the total mass in $V$ is $M=M(V)=\sum_{\mA}M_{\mA}(V)$. The constituent partial mass density $\tilde{\rho}_{\mA}$ and specific mass density $\rho_{\mA}>0$ are respectively defined as
\begin{subequations}\label{eq: def trhoA rhoA}
  \begin{align}
  \tilde{\rho}_{\mA}(\bx,t) :=&~ \displaystyle\lim_{ \vert V \vert \rightarrow 0} \dfrac{M_{\mA}(V)}{\vert V \vert},\\
  \rho_{\mA}(\bx,t) :=&~ \displaystyle\lim_{\vert V_{\mA}\vert  \rightarrow 0} \dfrac{M_{\mA}(V)}{\vert V_{\mA}\vert }.
\end{align}
\end{subequations}
The quantities represent the mass of the associated constituent $\mA$ per unit volume of the mixture $V$, and constituent volume $V_{\mA}$, respectively. In this paper we work with incompressible isothermal constituents of which the specific mass densities $\rho_{\mA}$ are constants. The density of the mixture is the sum of the partial mass densities of the constituents:
\begin{align}\label{eq: def rho}
\rho(\bx,t):=\displaystyle\sum_{\mA}\tilde{\rho}_{\mA}(\bx,t).
\end{align}
The volume fraction of constituent $\mA$ is defined as:
  \begin{align}\label{eq: def phi}
    \phi_{\mA}(\bx,t) :=&~ \displaystyle\lim_{\vert V \vert \rightarrow 0} \dfrac{\vert V_{\mA}\vert }{\vert V \vert}.
  \end{align}
We preclude the existence of void spaces by assuming:
\begin{align}\label{eq: sum phi}
    \displaystyle\sum_{\mA} \phi_{\mA}= 1.
\end{align}
The above definitions \eqref{eq: def trhoA rhoA}, \eqref{eq: def rho} and \eqref{eq: def phi} imply the relation:
\begin{align}\label{eq: relations rho and c}
  \tilde{\rho}_{\mA}(\bx,t) =&~ \rho_{\mA}\phi_{\mA}(\bx,t).
\end{align}
The constituent velocity is given by
\begin{align}
\bv_{\mA}(\mathbf{x},t)=\partial_t\bchi_{\mA}(\mathbf{X}_{\mA},t) \vert_{\mathbf{X}_\mA} =\grave{\mathbf{x}}_\mA (\mathbf{x},t),
\end{align}
where $\grave{\uppsi}$ is the time derivative of any differentiable function $\uppsi$ (of position and time) where the position $\mathbf{X}_\mA$ is fixed. Next, we denote the momentum of constituent $\mA$ as:
\begin{align}
    \mathbf{m}_{\mA}(\bx,t) = \trho_{\mA}(\bx,t) \bv_{\mA}(\bx,t).
\end{align}
By taking the sum of the momenta of the constituent we get the momentum of the mixture:
\begin{align}\label{eq: momentum mixture}
    \mathbf{m}(\bx,t) := \displaystyle\sum_{\mA} \mathbf{m}_{\mA}(\bx,t).
\end{align}
From the momentum of the mixture, we identify the \textit{mixture velocity} $\bv$ (also called mass-averaged velocity or barycentric velocity):
\begin{align}\label{eq: mix velo}
    \mathbf{m}(\bx,t) = \rho(\bx,t) \bv(\bx,t).
\end{align}
Another important velocity is the peculiar velocity (also known as diffusion velocity) of constituent $\mA$:
\begin{align}\label{eq: def bwj}
    \bw_{\mA}(\bx,t):=\bv_{\mA}(\bx,t)-\bv(\bx,t),
\end{align}
which describes the constituent velocity relative to the gross motion of the mixture. The peculiar velocity satisfies the property:
  \begin{align}\label{eq: rel gross motion zero}
    \displaystyle\sum_{\mA} \bJ_\mA  = \displaystyle\sum_{\mA} \rho_\mA^{-1} \bh_\mA  =&~ 0,
  \end{align}
where the so-called \textit{diffusive fluxes} are defined as:
\begin{subequations}\label{eq: def J and h}
\begin{align}
    \bh_\mA :=&~ \phi_\mA\bw_\mA ,\\
    \bJ_\mA :=&~ \trho_\mA \bw_\mA.   
\end{align}
\end{subequations}
Alongside the time derivative $\grave{\uppsi}$ of the differentiable function $\uppsi$ of $\mathbf{x}$ and $t$, we introduce a time derivative of $\uppsi$ that follows the mean motion. In the Eulerian frame these material derivatives are given by:
\begin{subequations}
\begin{align}
    \grave{\uppsi}=&~ \partial_t \uppsi + \bv_{\mA}\cdot \nabla \uppsi,\\
    \dot{\uppsi} =&~ \partial_t \uppsi + \bv\cdot \nabla \uppsi.\label{eq: mat der}
\end{align}
\end{subequations}

\subsection{Balance laws}\label{sec: BL}

According to the second metaphysical principle of the continuum theory of mixtures, the motion of each of the constituents is governed by an individual set of balance laws. These laws are contain interaction terms that model the interplay of the different constituents. Following e.g. \cite{truesdell1984historical}, each of the constituent $\mA = 1, \dots, N$ must satisfy in the following set of local balance laws for all $\mathbf{x} \in \Omega$ and $t \in (0,T)$:
\begin{subequations}\label{eq: BL const}
  \begin{align}
        \partial_t \tilde{\rho}_\mA + {\rm div}(\tilde{\rho}_\mA \bv_\mA) &=~ \gamma_\mA, \label{eq: local mass balance constituent j} \\
        \partial_t \mathbf{m}_\mA + {\rm div} \left( \mathbf{m}_\mA\otimes \bv_\mA \right) -  {\rm div} \mathbf{T}_\mA -  \trho_\mA \mathbf{b}_\mA &=~ \boldsymbol{\pi}_\mA,\label{eq: lin mom constituent j}\\
        \mathbf{T}_\mA-\mathbf{T}_\mA^T &=~\mathbf{N}_\mA,\label{eq: ang mom constituent j}\\
        \partial_t \left(\trho_\mA \left(\epsilon_\mA+\|\bv_\mA\|^2/2\right)\right) + \divg \left( \trho_\mA\left(\epsilon_\mA + \|\bv_\mA\|^2/2\right)\bv_\mA \right) &~\nn\\ 
        -\divg\left( \bv_\mA \mathbf{T}_\mA \right) - \trho_\mA \mathbf{b}_\mA\cdot\bv_\mA + \divg \mathbf{q}_\mA - \trho_\mA r_\mA &=~ e_\mA.\label{eq: energy constituent j}
  \end{align}
\end{subequations}
The equation \eqref{eq: local mass balance constituent j} represents the local constituent mass balance law, where the interaction term $\gamma_{\mA}$ is the mass supply of constituent $\mA$ due to chemical reactions with the other constituents. Next, \eqref{eq: lin mom constituent j} is the local constituent linear momentum balance law. Here $\mathbf{T}_\mA$ is the Cauchy stress tensor of constituent $\mA$, $\mathbf{b}_\mA$ the constituent external body force, and $\hat{\bpi}_\mA$ is the momentum exchange rate of constituent $\mA$ with the other constituents. In the remainder of the article we assume equal body forces ($\mathbf{b}_\mA= \mathbf{b}$ for $\mA = 1, \dots, N$). Moreover, we restrict to body forces of gravitational type: $\mathbf{b} = -b \boldsymbol{\jmath} = -b \nabla y$, with $y$ the vertical coordinate, $\boldsymbol{\jmath}$ the vertical unit vector and $b$ a constant. Next, \eqref{eq: ang mom constituent j} is the local constituent angular momentum balance with $\mathbf{N}_\mA$ the intrinsic moment of momentum. Finally, equation \eqref{eq: energy constituent j} is the local constituent energy balance. Here $\epsilon_\mA$ is the specific internal energy of constituent $\mA$, $\|\mathbf{v}_\mA\|=\sqrt{\mathbf{v}_\mA \cdot \mathbf{v}_\mA)}$ is the Euclidean norm of the velocity $\mathbf{v}_\mA$, $\mathbf{q}_\mA$ is the heat flux, $r_\mA$ is the external heat supply, and $e_\mA$ represents the energy exchange with the other constituents.

We denote the kinetic and gravitational energies of constituent respectively as:
\begin{subequations}
    \begin{align}
  \mathscr{K}_\mA =&~\trho_\mA \|\bv_\mA\|^2/2,\\
  \mathscr{G}_\mA =&~\trho_\mA b y.
\end{align}
\end{subequations}
On the account of the mass balance \eqref{eq: local mass balance constituent j} and the linear momentum balance \eqref{eq: lin mom constituent j}, we deduce the evolution of the constituent kinetic energy:
\begin{align}\label{eq: evo consti kin}
  \partial_t \mathscr{K}_\mA + {\rm div}\left( \mathscr{K}_\mA \mathbf{v}_\mA \right) - \mathbf{v}_\mA\cdot {\rm div} \mathbf{T}_\mA -  \trho_\mA \mathbf{b}_\mA \cdot \mathbf{v}_\mA=     \boldsymbol{\pi}_\mA\cdot\mathbf{v}_\mA - \frac{1}{2}\|\mathbf{v}_\mA\|^2 \gamma_\mA.
\end{align}
Next, the evolution of the gravitational energy follows from the constituent mass equation \eqref{eq: local mass balance constituent j}:
\begin{align}\label{eq: evo grav}
  \partial_t \mathscr{G}_\mA + {\rm div}\left( \mathscr{G}_\mA \mathbf{v}_\mA \right) + \trho_\mA \mathbf{v}_\mA\cdot \mathbf{b} - \gamma_\mA b y=0.
\end{align}
Taking the difference of \eqref{eq: energy constituent j} and \eqref{eq: evo consti kin} we obtain the evolution of the constituent internal energy:
\begin{align}\label{eq: internal energy constituent j}
    \partial_t \left(\trho_\mA \epsilon_\mA\right) + \divg \left( \trho_\mA\epsilon_\mA \bv_\mA \right) - \mathbf{T}_\mA: \nabla \mathbf{v}_\mA + \divg \mathbf{q}_\mA - \trho_\mA r_\mA  =\nn\\    - \boldsymbol{\pi}_\mA\cdot\mathbf{v}_\mA + \frac{1}{2}\|\mathbf{v}_\mA\|^2 \gamma_\mA + e_\mA .
\end{align}
The convective forms of the constituent evolution equations read:
\begin{subequations}\label{eq: BL const conv}
  \begin{align}
       \trho_\mA\grave{\bv}_\mA  +\tilde{\rho}_\mA {\rm div}\bv_\mA &=~ \gamma_\mA, \label{eq: local mass balance constituent j conv} \\
       \trho_\mA  \grave{\bv}_\mA - {\rm div} \mathbf{T}_\mA - \trho_\mA \mathbf{b}_\mA&=~  \mathbf{p}_\mA,\label{eq: lin mom constituent j conv}\\
        \trho_\mA \grave{\epsilon}_\mA -\mathbf{T}_\mA: \nabla \bv_\mA  + \divg \mathbf{q}_\mA -\trho_\mA r_\mA &=~  \breve{e}_\mA,\label{eq: energy constituent j conv}
  \end{align}
\end{subequations}
where the interaction terms are:
\begin{subequations}\label{eq: interaction conv}
\begin{align}
  \mathbf{p}_\mA =&~ \boldsymbol{\pi}_\mA - \gamma_\mA \bv_\mA, \\
  \breve{e}_\mA =&~ e_\mA - \boldsymbol{\pi}_\mA \cdot \mathbf{v}_\mA - \gamma_\mA (\epsilon_\mA - \|\mathbf{v}_\mA\|^2/2).
\end{align}
\end{subequations}
By invoking the constant specific densities $\rho_\mA$, we obtain the evolution equation of the volume fraction:
\begin{align}\label{eq: local mass balance constituent j phi}
    \partial_t \phi_\mA + {\rm div}(\phi_\mA \bv_\mA)= \dfrac{\gamma_\mA}{\rho_\mA}.
\end{align}

Next, we turn to the continuum balance laws of the mixtures. Summing the balance laws \eqref{eq: BL const} over the constituents gives:
\begin{subequations}\label{eq: BL mix}
  \begin{align}
        \partial_t \rho + {\rm div}(\rho \bv) &=~ 0, \label{eq: local mass balance mix} \\
        \partial_t \mathbf{m} + {\rm div} \left( \mathbf{m}\otimes \bv \right) -  {\rm div} \mathbf{T} -  \rho \mathbf{b} &=~0,\label{eq: lin mom mix}\\
        \mathbf{T}-\mathbf{T}^T &=~0,\label{eq: ang mom mix}\\
        \partial_t \left(\rho \left(\epsilon+\|\bv\|^2/2\right)\right) + \divg \left( \rho\left(\epsilon + \|\bv\|^2/2\right)\bv \right) &\nn\\ 
        -\divg\left( \mathbf{T}\bv\right)- \rho \mathbf{b}\cdot\bv +{\rm div}\mathbf{q}- \rho r &=~ 0.\label{eq: energy mix}
  \end{align}
\end{subequations}
where
\begin{subequations}
    \begin{align}
      \epsilon :=&~ \frac{1}{\rho}\displaystyle\sum_\mA \trho_\mA \left(\epsilon_\mA + \frac{1}{2} \|\mathbf{w}_\mA \|^2 \right),\\
       \mathbf{T} :=&~ \sum_\mA \mathbf{T}_\mA-\tilde{\rho}_\mA\bw_\mA\otimes\bw_\mA,\\
    \mathbf{b} :=&~\frac{1}{\rho}\sum_\mA \tilde{\rho}_\mA\mathbf{b}_\mA,\\
    \mathbf{q} :=&~ \displaystyle\sum_\mA \mathbf{q}_\mA - \mathbf{T}_\mA \mathbf{w}_\mA + \trho_\mA \left(\epsilon_\mA + \frac{1}{2}\|\mathbf{w}_\mA \|^2\right),\\
    r :=&~\frac{1}{\rho}\sum_\mA \tilde{\rho}_\mA r_\mA,   
    \end{align}
\end{subequations}
and where we have postulated the following balance conditions to hold:
\begin{subequations}
  \begin{align}
      \displaystyle\sum_\mA \gamma_\mA  =&~ 0,\label{eq: balance mass fluxes}\\
      \displaystyle\sum_\mA \boldsymbol{\pi}_\mA =&~ 0,\label{eq: balance momentum fluxes}\\
      \displaystyle\sum_\mA \mathbf{N}_\mA =&~ 0,\\
      \displaystyle\sum_\mA e_\mA =&~ 0.\label{eq: balance energy fluxes}
      \end{align}
\end{subequations}
In establishing the mixture laws \eqref{eq: BL mix} use has been made of the identities \eqref{eq: rel gross motion zero} and
\begin{align}
  \displaystyle\sum_\mA \trho_\mA \frac{1}{2}\|\mathbf{w}_\mA \|^2\mathbf{w}_\mA = 
  \displaystyle\sum_\mA \left(\trho_\mA \frac{1}{2}\|\mathbf{v}_\mA \|^2\mathbf{w}_\mA - 
  \trho_\mA \mathbf{w}_\mA (\mathbf{w}_\mA\cdot \mathbf{v})\right).
\end{align}

In agreement with the first metaphysical principle of mixture theory, the kinetic, gravitational and internal energy of the mixture are the superposition of the constituent energies:
\begin{subequations}
  \begin{align}
  \mathscr{K} =&~ \displaystyle\sum_{\mA} \mathscr{K}_\mA,\label{eq: def sum K}\\
  \mathscr{G} =&~ \displaystyle\sum_{\mA} \mathscr{G}_\mA,\\
  \mathscr{S} =&~ \sum_\mA \trho_\mA \epsilon_\mA.  
\end{align}
\end{subequations}
The kinetic energy of the mixture can be decomposed as:
\begin{subequations}\label{eq: relation kin energies}
    \begin{align}
      \mathscr{K} =&~ \bar{\mathscr{K}} + \displaystyle\sum_{\mA} \frac{1}{2} \trho_\mA \|\mathbf{w}_\mA\|^2,\\
       \bar{\mathscr{K}} =&~ \frac{1}{2} \rho \|\mathbf{v}\|^2,\label{eq: kin avg}
\end{align}
\end{subequations}
where $\bar{\mathscr{K}}$ is a kinetic energy of the mixture variables, and where the second term represents the kinetic energy of the constituents relative to the gross motion of the mixture. As a consequence, \eqref{eq: energy constituent j} represents the evolution of the internal and kinetic energy of the mixture
\begin{align}
  \partial_t \mathscr{E} + \divg \left( \mathscr{E}\bv \right) 
        -\divg\left( \bv\mathbf{T} \right)- \rho \mathbf{b}\cdot\bv +{\rm div}\mathbf{q}- \rho r&=~ 0,
\end{align}
with $\mathscr{E} = \mathscr{K}+\mathscr{G}+\mathscr{S}$, given the standing assumption of equal body forces. Finally, we remark that the system of mixture balance laws \eqref{eq: BL mix} may be augmented with evolution equations of the order parameters (mass and energy) and diffusive fluxes \cite{eikelder2023unified} to arrive at a system equivalent with \eqref{eq: BL const}.

\section{Constitutive modeling}\label{sec: 2nd law}
In this section we perform the constitutive modeling. We choose to employ the well-known Coleman-Noll procedure \cite{coleman1974thermodynamics} to construct constitutive models that satisfy the second law of thermodynamics. First, in \cref{sec: 2nd law mod} we introduce the second law of thermodynamics in the context of rational mechanics. Next, in \cref{sec: const mod res} we establish the constitutive modeling restriction yielding from the second law. Then, in \cref{sec: Selection of constitutive models} we select specific constitutive models compatible with the modeling restriction.

\subsection{Second law in mixture theory}\label{sec: 2nd law mod}

In agreement with the second metaphysical principle, the entropy of each of the constituents $\mA$ is governed by the balance law:
\begin{align}\label{eq: BL eta}
  \partial_t (\trho_\mA \eta_\mA) + {\rm div}\left( \trho_\mA \eta_\mA \mathbf{v}_\mA \right) + {\rm div} \left(\boldsymbol{\Phi}_\mA\right) - \trho_\mA s_\mA =\mathscr{P}_\mA,
\end{align}
where the constituent quantities are the specific entropy density $\eta_\mA$, the entropy flux $\boldsymbol{\Phi}_\mA$,  the specific entropy supply $s_\mA$, and the entropy production $\mathscr{P}_\mA$. The second law of thermodynamics dictates positive entropy production of the entire mixture:
\begin{align}\label{eq: second law}
 \displaystyle\sum_{\mA}  \mathscr{P}_\mA \geq 0.
\end{align}
The second law \eqref{eq: second law} is compatible with the first metaphysical principle of mixture theory.

In the following we derive the modeling restriction that results from the second law \eqref{eq: second law}. To this purpose, we introduce the \textit{Helmholtz mass-measure free energy} of constituent $\mA$:
\begin{align}\label{def: helmholtz constituent alpha}
    \psi_\mA := \epsilon_\mA - \theta \eta_\mA,
\end{align}
where $\theta$ is the temperature. We restrict to isothermal mixtures and thus all constituents have the same constant temperature $\theta = \theta_\mA$, $\mA = 1, \dots, N$. We now substitute \eqref{eq: BL eta} and \eqref{def: helmholtz constituent alpha} into \eqref{eq: second law} and arrive at:
\begin{align}\label{eq: second law form 2}
  \displaystyle\sum_{\mA} \partial_t (\trho_\mA \left(\epsilon_\mA-\psi_\mA\right)) + {\rm div}\left( \trho_\mA \left(\epsilon_\mA-\psi_\mA\right) \mathbf{v}_\mA \right)  + {\rm div} \left(\theta\boldsymbol{\Phi}_\mA\right) - \trho_\mA s_\mA \theta &~\geq 0.
\end{align}
We insert the balance of energy \eqref{eq: internal energy constituent j} into \eqref{eq: second law form 2} to arrive at:
\begin{align}\label{eq: balance entropy 3a}
   \displaystyle\sum_{\mA}    -\partial_t \left(\trho_\mA \psi_\mA\right) - \divg \left( \trho_\mA \psi_\mA \bv_\mA \right) +\mathbf{T}_\mA: \nabla \mathbf{v}_\mA+ \divg \left(\theta \bPhi_\mA-\mathbf{q}_\mA\right) &\nn\\
    + \trho_\mA \left(r_\mA -\theta s_\mA\right)  -\boldsymbol{\pi}_\mA\cdot\mathbf{v}_\mA+ \gamma_\mA \|\mathbf{v}_\mA\|^2/2&~\geq 0,
\end{align}
where the energy interaction term cancels because of \eqref{eq: balance energy fluxes}. In the final step we invoke the mass balance equation \eqref{eq: local mass balance constituent j} to find:
\begin{align}\label{eq: second law red 2}
\displaystyle\sum_{\mA} \trho_\mA\grave{\psi}_\mA - \mathbf{T}_\mA:\nabla \bv_\mA  + \divg\left(\mathbf{q}_\mA-\theta\boldsymbol{\Phi}_\mA\right)& \nn\\
+\trho_\mA \left(\theta s_\mA-r_\mA\right) +\boldsymbol{\pi}_\mA\cdot\mathbf{v}_\mA- \gamma_\mA \|\mathbf{v}_\mA\|^2/2 + \gamma_\mA \psi_\mA &~\leq 0.
\end{align}
This form of the second law provides the basis for the constitutive modeling.

Lastly, we remark that the second law may be written in an energy-dissipative form (given $r_\mA = \theta s_\mA$).
\begin{proposition}[Energy-dissipation]
The second law may be written as the energy-dissipation statement:
\begin{align}\label{eq: energy diss}
 \displaystyle\sum_\mA\left(\partial_t \mathscr{E}_\mA + {\rm div}\left(\mathscr{E}_\mA\bv_\mA \right) - {\rm div} \left(\mathbf{T}_\mA\bv_\mA   - \mathbf{q}_\mA+\theta\boldsymbol{\Phi}_\mA\right)\right)\leq 
 0,
\end{align}
with $\mathscr{E}_\mA = \mathscr{K}_\mA+\mathscr{G}_\mA+\trho_\mA \epsilon_\mA$, and where we have set $r_\mA = \theta s_\mA$.
\end{proposition}
\begin{proof}
Using the constituent mass equation \eqref{eq: local mass balance constituent j}, the second law \eqref{eq: second law red 2} may be written as:
\begin{align}\label{eq: second law alt}
&~\displaystyle\sum_{\mA}  \left[\partial_t (\trho_\mA\psi_\mA) + {\rm div}(\trho_\mA \psi_\mA \bv_\mA) - \mathbf{T}_\mA:\nabla \bv_\mA  + \divg\left(\mathbf{q}_\mA-\theta\boldsymbol{\Phi}_\mA\right) \right.\nn\\
&\quad\quad\quad\quad\quad \left.  +\boldsymbol{\pi}_\mA\cdot\mathbf{v}_\mA-e_\mA- \gamma_\mA \|\mathbf{v}_\mA\|^2/2 \right] \leq 0.
\end{align}
Adding \eqref{eq: evo consti kin} and \eqref{eq: evo grav} to the condition \eqref{eq: second law alt} provides the result.
\end{proof}
\subsection{Constitutive modeling restriction}\label{sec: const mod res}
We specify the modeling restriction \eqref{eq: second law red 2} to a particular set of constitutive constituent classes for the stress $\mathbf{T}_\mA$, free energy $\psi_\mA$, entropy flux $\bPhi_\mA$, momentum supply $\boldsymbol{\pi}_\mA$, and mass supply $\gamma_\mA$. We introduce the constitutive free energy class:
\begin{align}\label{eq: const classes free energy}
      \hat{\psi}_\mA =&~ \hat{\psi}_\mA(\phi_\mA,\nabla \phi_\mA, \mathbf{D}_\mA),
  \end{align}
and postpone the specification of the other constitutive classes. Here $\mathbf{D}_\mA$ is the symmetric velocity gradient of constituent $\mA$.


In the following we examine the constitutive modeling restriction \eqref{eq: second law red 2} for this specific set of constitutive classes. 
Substitution of the constitutive classes \eqref{eq: const classes free energy} into \eqref{eq: second law red 2} and expanding the peculiar derivative of the free energy provides:
\begin{align}\label{eq: second law red 3}
     \displaystyle\sum_\mA \trho_\mA\left(\dfrac{\partial \hat{\psi}_\mA}{\partial \phi_\mA}\grave{\phi}_\mA+\dfrac{\partial \hat{\psi}_\mA}{\partial \nabla\phi_\mA}\cdot\grave{\overline{\nabla\phi}}_\mA+\partial_{\mathbf{D}_\mA}\hat{\psi}_\mA\grave{\mathbf{D}}_\mA\right) - \hat{\mathbf{T}}_\mA:\nabla \bv_\mA &\nn\\
      + \divg \left(\mathbf{q}_\mA-\theta\hat{\bPhi}_\mA\right) +\trho_\mA \left(\theta s_\mA-r_\mA\right) &\nn\\
      +\boldsymbol{\pi}_\mA\cdot\mathbf{v}_\mA- \gamma_\mA \|\mathbf{v}_\mA\|^2/2 + \gamma_\mA \psi_\mA &~ \leq 0.
\end{align}
The arbitrariness of the peculiar time derivative $\grave{\mathbf{D}}_\mA$ precludes dependence of $\psi_\mA$ on $\mathbf{D}_\mA$. Thus, the free energy class reduces to:
\begin{align}\label{eq: free energy class psi}
      \hat{\psi}_\mA =&~\hat{\psi}_\mA (\phi_\mA,\nabla \phi_\mA),
\end{align}
and the last member in the first brackets is eliminated. 

Next we focus on the first term in the sum in \eqref{eq: second law red 3} 
and introduce the constituent quantity:
\begin{align}\label{eq: def mu}
   \chi_\mA = \phi_\mA \dfrac{ \partial \hat{\psi}_\mA}{\partial \phi_\mA} - \divg\left(\phi_\mA\dfrac{\partial \hat{\psi}_\mA}{\partial \nabla\phi_\mA}\right).
\end{align} 
\begin{lemma}[Identity peculiar derivative free energy]\label{lem: id free energy}
We have the identity:
  \begin{align}\label{eq: deduee 4}
\trho_\mA\left(\dfrac{\partial \hat{\psi}_\mA}{\partial \phi_\mA}\grave{\phi}_\mA+\dfrac{\partial \hat{\psi}_\mA}{\partial \nabla\phi_\mA}\cdot\grave{\overline{\nabla\phi}}_\mA\right) =&~-\trho_\mA\left(\chi_\mA\divg\bv_\mA+\left(\nabla \phi_\mA \otimes \dfrac{\partial \hat{\psi}_\mA}{\partial \nabla\phi_\mA}\right): \nabla\bv_\mA\right)\nn\\
&~-\divg\left(\trho_\mA\dfrac{\partial \hat{\psi}_\mA}{\partial \nabla\phi_\mA}\left(\phi_\mA\divg \bv_\mA\right)\right)\nn\\
&~+\gamma_\mA \chi_\mA + {\rm div}\left(\gamma_\mA\phi_\mA \dfrac{\partial \hat{\psi}_\mA}{\partial \nabla \phi_\mA}\right).  
\end{align}
\end{lemma}
\begin{proof}
Noting the identity
\begin{align}\label{eq: relation grad phi}
    \grave{\overline{\nabla \phi_\mA}} =&~ \nabla \left(\grave{\phi}_\mA\right) -(\nabla \phi_\mA)^T \nabla \bv_\mA,
\end{align} 
we can deduce:
\begin{align}\label{eq: deduee}
    \trho_\mA\dfrac{\partial \hat{\psi}_\mA}{\partial \nabla\phi_\mA}\cdot\grave{\overline{\nabla \phi_\mA}} =&~ \divg\left(\trho_\mA\dfrac{\partial \hat{\psi}_\mA}{\partial \nabla\phi_\mA}\grave{\phi}_\mA\right)-\grave{\phi}_\mA\divg \left( \trho_\mA\dfrac{\partial \hat{\psi}_\mA}{\partial \nabla\phi_\mA}\right)\nn\\
    &~-\trho_\mA\nabla \phi_\mA \otimes \dfrac{\partial \hat{\psi}_\mA}{\partial \nabla\phi_\mA}\cdot \nabla\bv_\mA.
\end{align}
By substituting the mass balance equation \eqref{eq: local mass balance constituent j} into \eqref{eq: deduee} we deduce:
\begin{align}\label{eq: deduee 2}
    \trho_\mA\dfrac{\partial \hat{\psi}_\mA}{\partial \nabla\phi_\mA}\grave{\overline{\nabla \phi_\mA}} =&~ -\divg\left(\trho_\mA\dfrac{\partial \hat{\psi}_\mA}{\partial \nabla\phi_\mA}\left(\phi_\mA\divg \bv_\mA-\rho_\mA^{-1}\gamma_\mA\right)\right)\nn\\
    &~+\left(\phi_\mA\divg \bv_\mA-\rho_\mA^{-1}\gamma_\mA\right)\divg \left( \trho_\mA\dfrac{\partial \hat{\psi}_\mA}{\partial \nabla\phi_\mA}\right)\nn\\
    &~-\left(\trho_\mA\nabla \phi_\mA \otimes \dfrac{\partial \hat{\psi}_\mA}{\partial \nabla\phi_\mA}\right): \nabla\bv_\mA.
\end{align}
As a result the first term in \eqref{eq: second law red 3} may be written as:
\begin{align}\label{eq: deduee 3}
&\trho_\mA\left(\dfrac{\partial \hat{\psi}_\mA}{\partial \phi_\mA}\grave{\phi}_\mA+\dfrac{\partial \hat{\psi}_\mA}{\partial \nabla\phi_\mA}\cdot\grave{\overline{\nabla\phi}}_\mA\right) = \nn\\
&~-\trho_\mA\left(\dfrac{\partial \hat{\psi}_\mA}{\partial \phi_\mA}\left(\phi_\mA\divg \bv_\mA-\rho_\mA^{-1}\gamma_\mA\right)\right)\nn\\
&~-\divg\left(\trho_\mA\dfrac{\partial \hat{\psi}_\mA}{\partial \nabla\phi_\mA}\left(\phi_\mA\divg \bv_\mA-\rho_\mA^{-1}\gamma_\mA\right)\right)\nn\\
&~+\left(\trho_\mA\divg \bv_\mA-\gamma_\mA\right)\divg \left( \phi_\mA\dfrac{\partial \hat{\psi}_\mA}{\partial \nabla\phi_\mA}\right)-\left(\trho_\mA\nabla \phi_\mA \otimes \dfrac{\partial \hat{\psi}_\mA}{\partial \nabla\phi_\mA}\right): \nabla\bv_\mA.
\end{align}
Substituting \eqref{eq: def mu} into \eqref{eq: deduee 3} completes the proof.
\end{proof}

Substitution of \cref{lem: id free energy} into the second law \eqref{eq: second law red 3} provides:
\begin{align}\label{eq: second law red 5}
     \displaystyle\sum_\mA -\left(\pi_\mA\mathbf{I}  + \trho_\mA \nabla \phi_\mA \otimes \dfrac{\partial \hat{\psi}_\mA}{\partial \nabla\phi_\mA}+\hat{\mathbf{T}}_\mA\right):\nabla \bv_\mA &\nn\\
     + \divg \left(\mathbf{q}_\mA-\theta\hat{\bPhi}_\mA-\dfrac{\partial \hat{\psi}_\mA}{\partial \nabla\phi_\mA}\phi_\mA\left(\trho_\mA \divg \bv_\mA-\gamma_\mA\right) \right)&\nn\\
     +\trho_\mA \left(\theta s_\mA-r_\mA\right)+\left(\boldsymbol{\pi}_\mA-\gamma_\mA\mathbf{v}_\mA/2\right)\cdot\mathbf{v}_\mA + \gamma_\mA\left( \psi_\mA + \chi_\mA\right)&~\leq 0,
\end{align}
where we have introduced $\pi_\mA := \trho_\mA \chi_\mA$.

At this point we remark that \eqref{eq: second law red 5} is degenerate because of the dependency of the various members in the superposition. Namely, the first two terms in the integral contain $\nabla \mathbf{v}_\mA$ and $\mathbf{v}_\mA$ are connected via the mass balance \eqref{eq: local mass balance constituent j}. 
To exploit the degeneracy, we introduce a scalar Lagrange multiplier $p \geq 0$ representing the \textit{mixture mechanical pressure}. Summation of \eqref{eq: local mass balance constituent j} over the constituents provides:
\begin{align}\label{eq: pressure eq}
    0=&~ p \displaystyle\sum_\mA \grave{\phi}_\mA +  \phi_\mA {\rm div}\bv_\mA - \rho_\mA^{-1}\gamma_\mA\nn\\
    =&~ p\displaystyle\sum_\mA  \bv_\mA \cdot \nabla \phi_\mA +    \phi_\mA {\rm div}\bv_\mA- \rho_\mA^{-1}\gamma_\mA,
\end{align}
where we recall the postulate of no excess volume \eqref{eq: sum phi}. Employing the relation \eqref{eq: pressure eq} into \eqref{eq: second law red 5} provides the requirement:
\begin{align}\label{eq: second law red 5 generate}
     \displaystyle\sum_\mA -\left((\pi_\mA+p\phi_\mA)\mathbf{I}  + \trho_\mA \nabla \phi_\mA \otimes \dfrac{\partial \hat{\psi}_\mA}{\partial \nabla\phi_\mA}+\hat{\mathbf{T}}_\mA\right):\nabla \bv_\mA &\nn\\
     + \divg \left(\mathbf{q}_\mA-\theta\hat{\bPhi}_\mA-\dfrac{\partial \hat{\psi}_\mA}{\partial \nabla\phi_\mA}\phi_\mA\left(\trho_\mA \divg \bv_\mA-\gamma_\mA\right)\right)
     +\trho_\mA \left(\theta s_\mA-r_\mA\right)&\nn\\     
     +\left(\boldsymbol{\pi}_\mA-\gamma_\mA\mathbf{v}_\mA/2-p\nabla \phi_\mA\right) \cdot\bv_\mA 
       + \gamma_\mA\left( \hat{\psi}_\mA + \chi_\mA+\rho_\mA^{-1}p\right) &~\leq 0.
\end{align}

The term $\mathfrak{p}_\mA := \pi_\mA+p\phi_\mA$ represents a generalized form of the constituent pressure in the incompressible mixture. It consists of the constituent mechanical pressure $p \phi_\mA$ and the constituent thermodynamical pressure $\pi_\mA$. 
The latter may be written in a form closely related to the classical thermodynamical pressure:
\begin{subequations}
    \begin{align}
  \pi_\mA =&~ \trho_\mA^2 \upsilon_\mA,\\
  \upsilon_\mA :=&~ \dfrac{ \partial \hat{\psi}_\mA}{\partial \trho_\mA} - \dfrac{1}{\trho_\mA}\divg\left(\trho_\mA\dfrac{\partial \hat{\psi}_\mA}{\partial \nabla\trho_\mA}\right).
\end{align}
\end{subequations}
Thus $\pi_\mA$ represents the thermodynamical pressure for the free energy constituent class \eqref{eq: free energy class psi}, where $\upsilon_\mA$ is a generalized derivative of the free energy.

We now introduce the volumetric Helmholtz free energy $\hat{\Psi}_\mA:=\trho_\mA\hat{\psi}_\mA$. Given the constituent class of $\hat{\psi}_\mA$ (equation \eqref{eq: free energy class psi}), we identify the volumetric Helmholtz free energy class:
\begin{align}\label{eq: free energy class Psi}
\hat{\Psi}_\mA=\hat{\Psi}_\mA(\phi_\mA,\nabla \phi_\mA)=\trho_\mA\hat{\psi}_\mA(\phi_\mA,\nabla \phi_\mA)=\rho_\mA\phi_\mA\hat{\psi}_\mA(\phi_\mA,\nabla \phi_\mA).
\end{align}
The constituent thermodynamical pressure $\pi_\mA$ may be written in terms of the volume-measure free energy $\hat{\Psi}_\mA$:
\begin{subequations}\label{eq: special form thermo pressure}
    \begin{align}
   \pi_\mA =&~ \phi_\mA \mu_\mA - \hat{\Psi}_\mA \label{eq: pi in terms op Upsilon}\\
    \mu_\mA
    :=&~ \dfrac{ \partial \hat{\Psi}_\mA}{\partial \phi_\mA} - \divg\left(\dfrac{\partial \hat{\Psi}_\mA}{\partial \nabla\phi_\mA}\right),
\end{align}
\end{subequations}
where $\mu_\mA$ is the chemical potential variable associated with the volume-measure free energy $\hat{\Psi}_\mA$. The volume-measure based chemical potential $\mu_\mA$ may be expressed in terms of the mass-measure based chemical potential $\tau_\mA$ via:
\begin{subequations}\label{eq: chemical potential upsilon}
    \begin{align}
    \mu_\mA =&~ \rho_\mA \left( \phi_\mA \tau_\mA + \hat{\psi}_\mA - \nabla \phi_\mA \cdot \dfrac{\partial \hat{\psi}_\mA}{\partial \nabla\phi_\mA}\right),\\
  \tau_\mA =&~ \dfrac{\partial \hat{\psi}_\mA}{\partial\phi_\mA} - {\rm div}\left( \dfrac{\partial \hat{\psi}_\mA}{\partial \nabla \phi_\mA} \right).
\end{align}
\end{subequations}

\begin{remark}[Dalton's law]\label{rmk: Dalton}
   The mechanical pressure obeys Dalton's law. Namely, the constituent mechanical pressure $p \phi_\mA$ is the product of the mixture mechanical pressure $p$ and the constituent volume fraction $\phi_\mA$. Additionally, according to the axiom \eqref{eq: sum phi}, the sum of the constituent mechanical pressures is the mixture mechanical pressure $p$.
\end{remark}
\begin{remark}[Incompressibility constraint]
The introduction of the mixture mechanical pressure is connected with an incompressibility constraint in absense of mass fluxes (i.e. $\gamma_\mA = 0$). Namely, by introducing the mean velocity \begin{align}\label{eq: volume-averaged velocity}
    \bu := \displaystyle\sum_\mA \phi_\mA \bv_\mA,
\end{align}
\eqref{eq: pressure eq} takes the form: 
\begin{align}
     p {\rm div} \bu = p \displaystyle\sum_\mA {\rm div}(\phi_\mA\bv_\mA) = p \displaystyle\sum_\mA \bv_\mA \cdot \nabla \phi_\mA +  \phi_\mA {\rm div}\bv_\mA =0,
\end{align}
provided $\gamma_\mA = 0$.
The mean velocity $\bu$ is known as the volume averaged velocity which is an incompressible field in absense of mass fluxes. The observation has been employed in the formulation of reduced (approximate) quasi-incompressible Navier-Stokes Cahn-Hilliard models \cite{boyer2002theoretical,ding2007diffuse,abels2012thermodynamically,eikelder2023unified} with an incompressible velocity field. 
\end{remark}

Based on the condition \eqref{eq: second law red 5 generate}, we restrict to the  following constitutive constituent classes for the stress $\mathbf{T}_\mA$, entropy flux $\bPhi_\mA$, entropy supply $s_\mA$, mass supply $\gamma_\mA$, and momentum supply $\boldsymbol{\pi}_\mA$:
\begin{subequations}\label{eq: const classes}
  \begin{align}
      \hat{\bPhi}_\mA =&~ \hat{\bPhi}_\mA\left(\phi_\mA,\nabla \phi_\mA, {\rm div}\mathbf{v}_\mA, \mathbf{q}_\mA, \gamma_\mA\right),\\
      \hat{s}_\mA =&~ \hat{s}_\mA\left(r_\mA\right),\\
      \hat{\mathbf{T}}_\mA =&~ \hat{\mathbf{T}}_\mA(\phi_\mA,\nabla \phi_\mA, \mathbf{D}_\mA, \pi_\mA, p),\\
      \hat{\gamma}_\mA =&~ \hat{\gamma}_\mA\left(\phi_\mA,\nabla \phi_\mA, p,\left\{\psi_\mB\right\}_{\mB=1,\dots, N},\left\{\mu_\mB\right\}_{\mB=1,\dots, N}\right),\label{eq: const classes gamma}\\
      \hat{\bpi}_\mA =&~ \hat{\bpi}_\mA\left(\phi_\mA,\nabla \phi_\mA, \left\{\mathbf{v}_\mB\right\}_{\mB=1,\dots, N}, \left\{\gamma_\mB\right\}_{\mB=1,\dots, N}\right),\label{eq: const classes bpm}
  \end{align}
\end{subequations}
where in \eqref{eq: const classes gamma} and \eqref{eq: const classes bpm} the dependence on the sets over all constituents is a consequence of the axioms \eqref{eq: balance mass fluxes} and \eqref{eq: balance momentum fluxes}.

\subsection{Selection of constitutive models}\label{sec: Selection of constitutive models}
We are now in the position to pose thermodynamically consistent relations for the constitutive classes \eqref{eq: const classes}.\\

\noindent \textit{Entropy flux}. By demanding the divergence term to equate zero, we identify the entropy flux of constituent $\mA$ as:
\begin{align}\label{eq: entropy flux expression}
    \hat{\bPhi}_\mA \equiv \dfrac{\mathbf{q}_\mA}{\theta}-\dfrac{1}{\theta}\dfrac{\partial \hat{\psi}_\mA}{\partial \nabla \phi_\mA}\phi_\mA\left(\trho_\mA \divg \bv_\mA-\hat{\gamma}_\mA\right).
\end{align}
The first member in the entropy flux is the constituent version of the classical term that appears in single constituent models. On the other hand, the second member in the entropy flux is the incompressible counterpart augmented with mass transfer, of the so-called \textit{extra entropy flux}.\\

\noindent \textit{Entropy supply}. By requiring the last member in \eqref{eq: second law red 5 generate} to disappear, we identify the constituent entropy supply density as:
\begin{align}\label{eq: entropy supply density expression}
    s_\mA \equiv \dfrac{r_\mA}{\theta}. 
\end{align}

\noindent \textit{Stress tensor}. To preclude that variations of the velocity gradient $\nabla \bv_\mA$ cause a violation of the second law \eqref{eq: second law red 5 generate} we insist:
\begin{align}\label{eq: req stress}
-\left((\trho_\mA\chi_\mA+p\phi_\mA)\mathbf{I}  + \trho_\mA \nabla \phi_\mA \otimes \dfrac{\partial \hat{\psi}_\mA}{\partial \nabla \phi_\mA}+\hat{\mathbf{T}}_\mA\right):\nabla \bv_\mA \leq 0.
\end{align}
We select the following constitutive model for the stress tensor that is compatible with \eqref{eq: req stress}:
  \begin{align}\label{eq: stress expressions}
      \hat{\mathbf{T}}_\mA =&~ \tilde{\nu}_\mA \left(2\mathbf{D}_\mA + \lambda_\mA ({\rm div}\mathbf{v}_\mA)\mathbf{I}\right) - (\pi_\mA+p\phi_\mA) \mathbf{I} - \trho_\mA\nabla \phi_\mA \otimes \dfrac{\partial \hat{\psi}_\mA}{\partial \nabla\phi_\mA},
  \end{align}
where $\tilde{\nu}_\mA = \nu_\mA \phi_\mA \geq 0$ is a dynamic viscosity, and $\lambda_\mA \geq -2/d$.
\begin{lemma}[Compatibility stress tensor]\label{lem: compatibility stress tensor}
The choice \eqref{eq: stress expressions} is compatible with the thermodynamical restriction \eqref{eq: req stress}.
\end{lemma}
\begin{proof}
  This is a standard result. In this particular case \eqref{eq: req stress} takes the form:
  \begin{align}\label{eq: second law viscosity}
    -2 \tilde{\nu}_\mA \left( \mathbf{D} - \frac{1}{d} ({\rm div} \mathbf{v}_\mA) \mathbf{I}\right):\left(\mathbf{D} - \frac{1}{d} ({\rm div} \mathbf{v}_\mA) \mathbf{I}\right) - \tilde{\nu}_\mA\left(\lambda_\mA + \frac{2}{d}\right)\left({\rm div} \mathbf{v}_\mA\right)^2 \leq 0.
  \end{align}
\end{proof}
\begin{remark}[General form stress tensor]
  The requirement \eqref{eq: req stress} implies the general form:
    \begin{align}
      \hat{\mathbf{T}}_\mA =&~ 2 \mathbf{K}_\mA \mathbf{D}_\mA - (\pi_\mA+p\phi_\mA) \mathbf{I} - \trho_\mA\nabla \phi_\mA \otimes \dfrac{\partial \hat{\psi}_\mA}{\partial \nabla\phi_\mA},
  \end{align}
  where $\mathbf{K}_\mA = \mathbf{K}_\mA(\phi_\mA,\nabla \phi_\mA, \mathbf{D}_\mA)$ is a quantity that satisfies:
    \begin{align}
      \mathbf{D}_\mA^T\mathbf{K}_\mA\mathbf{D}_\mA  \geq 0.
  \end{align}
  This implication follows from a result concerning thermodynamical inequalities proved by Gurtin \cite{gurtin1996generalized}. 
\end{remark}

\noindent \textit{Mass transfer}.
To rule out violations \eqref{eq: second law red 5 generate} caused by the latter term on the left-hand side, we impose the following requirement on the mass interaction terms:
    \begin{align}\label{eq: req gamma}
   \displaystyle\sum_\mA \hat{\gamma}_\mA\left( \psi_\mA + \chi_\mA+\rho_\mA^{-1}p\right) &~\leq 0.
\end{align}
The requirement distinguishes from the compressible situation by the occurrence of the hydrodynamic pressure $p$, see e.g. Morro \cite{morro2016nonlinear}. We take the following model for the mass transfer:
\begin{subequations}\label{eq: choice gamma}
    \begin{align}
  \hat{\gamma}_\mA =&~ -\hat{m}_\mA \left( (\psi_\mA-\psi_N) + (\chi_\mA-\chi_N)+(\rho_\mA^{-1}-\rho_N^{-1})p\right), \label{eq: choice gamma mA} \nn\\
  & \quad \quad \quad \quad \quad \quad \quad \quad \quad \text{ for }\mA = 1, \dots, N-1,\\  
  \hat{\gamma}_N =&~ - \displaystyle\sum_{\mA=1, \dots, N-1} \hat{\gamma}_\mA, \label{eq: choice gamma N}
\end{align}
\end{subequations}
for some non-negative constituent quantity $\hat{m}_\mA \geq 0$ that vanishes when $\phi_\mA = 0,1$.

\begin{lemma}[Compatibility mass transfer]\label{lem: compatibility gamma}
The choice \eqref{eq: choice gamma} is compatible with the balance of mass supply \eqref{eq: balance momentum fluxes}, and the thermodynamical restriction \eqref{eq: req gamma}.
\end{lemma}
\begin{proof}
Invoking the identity \eqref{eq: balance momentum fluxes} written as \eqref{eq: choice gamma N}, the condition \eqref{eq: req gamma} is equivalent to:
\begin{align}\label{eq: req pi 2}
   \displaystyle\sum_{\mA=1,\dots,N-1} \hat{\gamma}_\mA\left( (\psi_\mA-\psi_N) + (\chi_\mA-\chi_N)+(\rho_\mA^{-1}-\rho_N^{-1})p\right) \leq 0.
\end{align}
The choice \eqref{eq: choice gamma} causes each of the terms in the sum in \eqref{eq: req pi 2} to be non-positive. Compatibility with \eqref{eq: balance momentum fluxes} follows from \eqref{eq: choice gamma N}.
\end{proof}

On the account of the identity:
\begin{align}
  \rho_\mA \left(\psi_\mA + \chi_\mA\right) = \mu_\mA,
\end{align}
the mass flux may be expressed in terms of the chemical potential $\mu_\mA$:
\begin{align}\label{eq: compact gamma}
  \hat{\gamma}_\mA =&~ -\hat{m}_\mA \left( \frac{1}{\rho_\mA}\left(\mu_\mA+p\right) - \frac{1}{\rho_N}\left(\mu_N+p\right)\right),\quad \text{ for }\mA = 1, \dots, N-1.
\end{align}
Furthermore, the mass flux may be written as:
  \begin{align}
    \hat{\gamma}_\mA =&~ - \hat{m}_\mA (g_\mA - g_N),  \quad \quad \text{ for }\mA = 1, \dots, N-1.
  \end{align}
where $g_\mA$ represents the Gibbs free energy of constituent $\mA$:
\begin{align}
    g_\mA = \psi_\mA + \frac{\mathfrak{p}_\mA}{\trho_\mA} =\psi_\mA + \chi_\mA + \frac{p}{\rho_\mA},
\end{align}
and where we recall the total constituent pressure $\mathfrak{p}_\mA = \pi_\mA + \phi_\mA p$.\\

\noindent \textit{Momentum transfer}.
To avoid a violation of \eqref{eq: second law red 5 generate} resulting from momentum transfer, we demand:
    \begin{align}\label{eq: req pi}
   \displaystyle\sum_\mA \bv_\mA \cdot (\bpi_\mA-\hat{\gamma}_\mA\mathbf{v}_\mA/2-p\nabla \phi_\mA) &~\leq 0.
\end{align}
We select the mass transfer model:
\begin{align}\label{eq: choice pi}
  \bpi_\mA = p \nabla \phi_\mA+\displaystyle\sum_\mB R_{\mA\mB}(\bw_\mB-\bw_\mA) + \boldsymbol{\beta}_\mA,
\end{align}
where
\begin{subequations}\label{eq: def beta}
    \begin{align}
  \boldsymbol{\beta}_\mA =&~ \frac{1}{2} \hat{\gamma}_\mA \left(\mathbf{w}_\mA +\mathbf{w}_N + 2 \mathbf{v}\right),  \quad \text{ for }\mA = 1, \dots, N-1, \label{eq: def beta mA} \\
  \boldsymbol{\beta}_N =&~ -\displaystyle\sum_{\mA = 1, \dots, N-1} \boldsymbol{\beta}_\mA. \label{eq: def beta N}
\end{align}
\end{subequations}
Furthermore, $R_{\mA\mB}$ is a symmetric non-negative matrix of the form:
\begin{align}\label{eq: R}
    R_{\mA\mB} = \dfrac{p\phi_\mA\phi_\mB}{D_{\mA\mB}} \geq 0,
\end{align}
with $D_{\mA\mB} \geq 0$ a symmetric diffusion coefficient. 
\begin{lemma}[Compatibility momentum transfer]\label{lem: compatibility pi}
The momentum transfer model \eqref{eq: choice pi} is compatible with the balance of momentum supply \eqref{eq: balance momentum fluxes}, and the thermodynamical restriction \eqref{eq: req pi}.
\end{lemma}
\begin{proof}
Compatibility with \eqref{eq: balance momentum fluxes} is a consequence of \eqref{eq: choice pi}, the symmetry of $R_{\mA\mB}$, and the definition \eqref{eq: def beta}. Next, recalling the axiom of constant volume \eqref{eq: sum phi}, the axioms of balance of mixture mass and momentum \eqref{eq: balance mass fluxes}-\eqref{eq: balance momentum fluxes}, the condition \eqref{eq: req pi} is equivalent to:
\begin{align}\label{eq: req pi 2a}
    \displaystyle\sum_\mA \bw_\mA \cdot \left(\bpi_\mA - p \nabla \phi_\mA - \hat{\gamma}_\mA \left( \frac{1}{2}\mathbf{w}_\mA + \mathbf{v} \right)\right) \leq 0.
\end{align}
Substitution of \eqref{eq: choice pi} into \eqref{eq: req pi 2a} provides the requirement:
\begin{align}\label{eq: req pi 3}
    \displaystyle\sum_{\mA,\mB} R_{\mA\mB}\bw_\mA \cdot(\bw_\mB-\bw_\mA) + \displaystyle\sum_\mA \bw_\mA \cdot\left(\boldsymbol{\beta}_\mA- \hat{\gamma}_\mA  \left( \frac{1}{2}\mathbf{w}_\mA + \mathbf{v} \right)\right) \leq 0.
\end{align}
The first term is non-positive as a consequence of the identity:
\begin{align}
    \displaystyle\sum_{\mA,\mB} R_{\mA\mB} (\bw_\mB-\bw_\mA)\cdot \bw_\mA =  -\frac{1}{2}\displaystyle\sum_{\mA,\mB} R_{\mA\mB} \|\bw_\mA-\bw_\mB\|^2.
\end{align}
Taking the second term in isolation, splitting the summation provides:
\begin{align}\label{eq: req pi 3b}
    &\displaystyle\sum_\mA \bw_\mA \cdot\left(\boldsymbol{\beta}_\mA- \hat{\gamma}_\mA  \left( \frac{1}{2}\mathbf{w}_\mA + \mathbf{v} \right)\right) =\nn\\
    &\quad \quad \quad \quad \displaystyle\sum_{\mA=1,\dots N-1} \bw_\mA \cdot\left(\boldsymbol{\beta}_\mA- \hat{\gamma}_\mA  \left( \frac{1}{2}\mathbf{w}_\mA + \mathbf{v} \right)\right) \nn\\
    &\quad \quad \quad \quad + \bw_N \cdot\left(\boldsymbol{\beta}_N- \hat{\gamma}_N  \left( \frac{1}{2}\mathbf{w}_N + \mathbf{v} \right)\right).
\end{align}
We substitute the identities \eqref{eq: choice gamma N} and \eqref{eq: def beta N} arrive at:
\begin{align}\label{eq: req pi 3c}
    &\displaystyle\sum_\mA \bw_\mA \cdot\left(\boldsymbol{\beta}_\mA- \hat{\gamma}_\mA  \left( \frac{1}{2}\mathbf{w}_\mA + \mathbf{v} \right)\right) =\nn\\
    &\displaystyle\sum_{\mA=1, \dots, N-1} (\bw_\mA-\bw_N)\cdot \left(\boldsymbol{\beta}_\mA- \frac{1}{2}\hat{\gamma}_\mA \left(\mathbf{w}_\mA +\mathbf{w}_N\right) - \hat{\gamma}_\mA \mathbf{v}\right).
\end{align}
Inserting the definition \eqref{eq: def beta mA} causes the term to vanish.
\end{proof}

\begin{remark}[Stefan-Maxwell model]
The second member in \eqref{eq: choice pi} represents an \textit{isothermal Stefan-Maxwell model} \cite{whitaker2009derivation}. The term $p \phi_\mA\phi_\mB$ is proportional to the frequency of collisions between $\mA$ and $\mB$. This makes intuitive sense in the way that the force that is exerted by constituent $\mB$ on constituent $\mA$ scales with the frequency of collisions between the two constituents. Provided mass transfer is absent ($\hat{\gamma}_\mA = 0$), the momentum transfer vanishes if and only if:
\begin{align}\label{eq: Stefan-Maxwell equations}
    \nabla \phi_\mA  +\displaystyle\sum_\mB \dfrac{\phi_\mA\phi_\mB}{D_{\mA\mB}}(\bv_\mB-\bv_\mA)=0.
\end{align}
The equations \eqref{eq: Stefan-Maxwell equations} represent the well-known \textit{Stefan-Maxwell equations} that describe an equilibrium situation. The left-hand side of \eqref{eq: Stefan-Maxwell equations} represents the diffusion driving force for constituent $\mA$, whereas the right-hand side of \eqref{eq: Stefan-Maxwell equations} is the drag force on constituent $\mA$ that resists the diffusion. As such $D_{\mA\mB}$ can be interpreted as an inverse drag coefficient, and is referred to as \textit{Stefan-Maxwell diffusivity}.
\end{remark}

This concludes the Coleman-Noll procedure. We have now obtained the \textit{incompressible multi-constituent model} that is consistent with the second law of mixture-theory:
\begin{subequations}\label{eq: model sum}
  \begin{align}
 \partial_t \trho_\mA + {\rm div}(\trho_\mA \bv_\mA) - \hat{\gamma}_\mA &=~ 0, \label{eq: model sum: cont}\\
 \partial_t (\trho_\mA \bv_\mA) + {\rm div} \left( \trho_\mA \bv_\mA\otimes \bv_\mA \right)  + \phi_\mA\nabla p  \nn\\
 - {\rm div}\left(\tilde{\nu}_\mA \left(2\mathbf{D}_\mA + \lambda_\mA {\rm div}\mathbf{v}_\mA\right)\right) &\nn\\
 + \nabla \pi_\mA + {\rm div} \left(\trho_\mA \nabla \phi_\mA \otimes \dfrac{\partial \hat{\psi}_\mA}{\partial \nabla\phi_\mA} \right)-\trho_\mA\mathbf{b}&\nn\\
    -\displaystyle\sum_\mB \dfrac{p\phi_\mA\phi_\mB}{D_{\mA\mB}}(\bv_\mB-\bv_\mA)- \boldsymbol{\beta}_\mA&=~ 0, \label{eq: model sum: mom}
  \end{align}
\end{subequations}
for $\mA = 1,...,N$ where $\hat{\gamma}_\mA$ and $\boldsymbol{\beta}_\mA$ are given in \eqref{eq: choice gamma} and \eqref{eq: def beta}, respectively.

We now discuss some properties of the model. First we explicitly state the compatibility with the second law.
\begin{theorem}[Compatibility second law]\label{eq: compatibility 2nd law}
The model \eqref{eq: model sum} is compatible with the second law of  thermodynamics \eqref{eq: second law}.
\end{theorem}
\begin{proof}
  This follows from the form of the second law \eqref{eq: second law red 5 generate} and \cref{lem: compatibility stress tensor}, \cref{lem: compatibility gamma}, and \cref{lem: compatibility pi}. In particular, inserting \eqref{eq: entropy flux expression}, \eqref{eq: entropy supply density expression}, \eqref{eq: stress expressions}, \eqref{eq: choice gamma} and \eqref{eq: choice pi} into \eqref{eq: second law red 5 generate} reveals that the second law is satisfied with
\begin{align}\label{eq: second law red 7}
    \theta \displaystyle\sum_{\mA}  \mathscr{P}_\mA = &~ \displaystyle\sum_\mA 2 \tilde{\nu}_\mA \left( \mathbf{D} - \frac{1}{d} ({\rm div} \mathbf{v}_\mA) \mathbf{I}\right):\left(\mathbf{D} - \frac{1}{d} ({\rm div} \mathbf{v}_\mA) \mathbf{I}\right)\nn\\
    &~+ \displaystyle\sum_\mA \tilde{\nu}_\mA\left(\lambda_\mA + \frac{2}{d}\right)\left({\rm div} \mathbf{v}_\mA\right)^2 + \frac{1}{2}\displaystyle\sum_{\mA,\mB} R_{\mA\mB}\|\bw_\mA-\bw_\mB\|^2 \nn\\
    &~+\displaystyle\sum_{\mA = 1, \dots, N-1} \hat{m}_\mA \left(g_\mA - g_N\right)^2\geq 0.
\end{align}
\end{proof}

We now note the reduction to the standard Navier-Stokes equations in the single fluid regime.
\begin{proposition}[Reduction to Navier-Stokes]
The multi-constituent system \eqref{eq: model sum} reduces to the standard incompressible Navier-Stokes equations in the single-constituent regime ($\phi_\mA = 1$):
\begin{subequations}\label{eq: NS}
  \begin{align} 
 \partial_t (\rho_\mA \bv_\mA) + {\rm div} \left( \rho_\mA \bv_\mA\otimes \bv_\mA \right)  + \nabla p &\nn\\
 - {\rm div} \left(\nu_\mA \left(2\mathbf{D}_\mA + \lambda_\mA {\rm div}\mathbf{v}_\mA\right) \right) -\rho_\mA\mathbf{b}&=~ 0, \\
  {\rm div}\bv_\mA &=~ 0,
  \end{align}
\end{subequations}
with $\rho_\mA = \rho, \mathbf{v}_\mA = \mathbf{v}$, and $\mathbf{D}_\mA = \mathbf{D}:=(\nabla \mathbf{v} + (\nabla \mathbf{v})^T)/2$.
\end{proposition}

We finalize this section with a more compact form of the mixture model.
\begin{lemma}[Compact form free energy contributions]\label{lem: free energy contributions}
The free energy contributions in the momentum equation may be expressed in the compact form:
\begin{align}
\phi_\mA \nabla \mu_\mA=\nabla \pi_\mA + {\rm div} \left(\trho_\mA \nabla \phi_\mA \otimes \dfrac{\partial \hat{\psi}_\mA}{\partial \nabla\phi_\mA} \right).
\end{align}
\end{lemma}
\begin{proof}
Substituting \eqref{eq: special form thermo pressure} and subsequently expanding the derivatives yields:
\begin{align}\label{eq: final expression}
&\nabla \pi_\mA + {\rm div} \left(\trho_\mA \nabla \phi_\mA \otimes \dfrac{\partial \hat{\psi}_\mA}{\partial \nabla\phi_\mA} \right) =\nn\\
&\nabla \left(\phi_\mA \mu_\mA - \hat{\Psi}_\mA\right) + {\rm div} \left(\nabla \phi_\mA \otimes \dfrac{\partial \hat{\Psi}_\mA}{\partial \nabla\phi_\mA} \right) =\nn\\
&\phi_\mA \nabla\mu_\mA + \nabla \phi_\mA \dfrac{ \partial \hat{\Psi}_\mA}{\partial \phi_\mA} - \nabla \phi_\mA \divg\left(\dfrac{\partial \hat{\Psi}_\mA}{\partial \nabla\phi_\mA}\right)  - \nabla \hat{\Psi}_\mA \nn\\
&+ \nabla \phi_\mA {\rm div} \left( \dfrac{\partial \hat{\Psi}_\mA}{\partial \nabla\phi_\mA} \right)+\left(\mathbf{H} \phi_\mA\right) \dfrac{\partial \hat{\Psi}_\mA}{\partial \nabla\phi_\mA} =\nn\\
&\phi_\mA \nabla\mu_\mA - \nabla \hat{\Psi}_\mA +\nabla \phi_\mA \dfrac{ \partial \hat{\Psi}_\mA}{\partial \phi_\mA}   +\left(\mathbf{H} \phi_\mA\right) \dfrac{\partial \hat{\Psi}_\mA}{\partial \nabla\phi_\mA},
\end{align}
where $\mathbf{H} \phi_\mA$ is the hessian of $\phi_\mA$.
As a consequence of the volumetric Helmholtz free energy class \eqref{eq: free energy class Psi}, the latter three terms in the final expression in \eqref{eq: final expression} vanish.
\end{proof}

On the account of \cref{lem: free energy contributions}, the multi-constituent model \eqref{eq: model sum} takes the more compact form:
\begin{subequations}\label{eq: model FE}
  \begin{align}
 \partial_t \trho_\mA + {\rm div}(\trho_\mA \bv_\mA) -\hat{\gamma}_\mA &=~ 0, \label{eq: model FE: cont}\\
 \partial_t (\trho_\mA \bv_\mA) + {\rm div} \left( \trho_\mA \bv_\mA\otimes \bv_\mA \right)  + \phi_\mA\nabla \left(p + \mu_\mA \right)  \nn\\
 - {\rm div}\left(\tilde{\nu}_\mA \left(2\mathbf{D}_\mA + \lambda_\mA {\rm div}\mathbf{v}_\mA\right)\right) -\trho_\mA\mathbf{b}&\nn\\
    -\displaystyle\sum_\mB \dfrac{p\phi_\mA\phi_\mB}{D_{\mA\mB}}(\bv_\mB-\bv_\mA)- \boldsymbol{\beta}_\mA&=~ 0, \label{eq: model FE: mom}
  \end{align}
\end{subequations}
for $\mA = 1,...,N$.


\section{Diffuse-interface models}\label{sec: diffuse interface}

In this section we present diffuse-interface models. First, in \cref{subsec: GL Free Energy} we introduce the Ginzburg-Landau free energy. Next, in \cref{subsec: dim less} we provide the dimensionless form of the model. Finally, in \cref{subsec: equilibrium prof} we discuss the equilibrium profile of the mixture model. 

\subsection{Ginzburg-Landau free energy}\label{subsec: GL Free Energy}
Important classes of fluid mixture models arise when selecting the constituent Helmholtz free energy to be of Ginzburg-Landau type. We consider two different options: (I) a Ginzburg-Landau type volume-measure-based free energy, and (II) a Ginzburg-Landau type volume-measure-based free energy.\\

\noindent \textit{Model I}. The Helmholtz volume-measure free energy is given by:
\begin{subequations}\label{eq: selection GL form Psi}
\begin{align}
    \hat{\Psi}_\mA^{\rm I} =&~ \dfrac{\sigma_{\mA}}{\varepsilon_{\mA}}W(\phi_\mA) + \sigma_{\mA}\varepsilon_{\mA}\|\nabla \phi_\mA\|^2\\
    W(\phi_\mA)=&~2\phi^2_\mA(1-\phi_\mA)^2,\label{eq: selection GL form Psi W} 
\end{align}
\end{subequations}
where $W=W(\phi_\mA)$ represents a double-well potential, $\varepsilon_{\mA}$ are interface thickness variables, and $\sigma_{\mA}$ are quantities related to the surface energy density. We assume that $\varepsilon_{\mA}$ and $\sigma_\mA$ are constants. The chemical potential takes the form:
    \begin{align}\label{eq: CP in form I GL}
  \mu^{\rm I}_\mA =&~   \dfrac{\sigma_{\mA}}{\varepsilon_{\mA}}W'(\phi_\mA) -2 \sigma_{\mA}\varepsilon_{\mA}\Delta \phi_\mA,
\end{align}
Furthermore, the mass flux takes the form:
\begin{align}\label{eq: gamma GL form I}
  \hat{\gamma}_\mA^{\rm I} =&~ -\hat{m}_\mA \left( \dfrac{\sigma_{\mA}}{\rho_\mA\varepsilon_{\mA}}W'(\phi_\mA)- \dfrac{\sigma_{N}}{\rho_N\varepsilon_{N}}W'(\phi_N)\right.\nn\\
  &\quad\quad\quad\left.-2 \frac{\sigma_{\mA}}{\rho_\mA}\varepsilon_{\mA}\Delta \phi_\mA  +2 \frac{\sigma_N}{\rho_N}\varepsilon_N\Delta \phi_N + \left(\frac{1}{\rho_\mA}-\frac{1}{\rho_N}\right)p\right),
\end{align}
for $\mA = 1, \dots, N-1$ and \eqref{eq: choice gamma N} for $\mA = N$.\\

\noindent \textit{Model II}. 
The Helmholtz mass-measure free energy reads:
\begin{align}\label{eq: selection GL form psi}
    \hat{\psi}_\mA^{\rm II} =&~ 2\dfrac{\kappa_{\mA}}{\varepsilon_{\mA}}W(\phi_\mA) + 2\kappa_{\mA}\varepsilon_{\mA}\|\nabla \phi_\mA\|^2,
\end{align}
where $W=W(\phi_\mA)$ is given in \eqref{eq: selection GL form Psi W}. Also in this second model, the interface thickness variables $\varepsilon_{\mA}$ and surface energy density quantities $\kappa_\mA$ are assumed constant. The associated chemical potential takes the form:
    \begin{align}\label{eq: CP in form II GL}
  \tau^{\rm II}_\mA= &~  2\dfrac{\kappa_{\mA}}{\varepsilon_{\mA}}W'(\phi_\mA) -4 \kappa_{\mA}\varepsilon_{\mA}\Delta \phi_\mA,
\end{align} 
The corresponding mass flux reads:
\begin{align}\label{eq: gamma GL  form II}
  \hat{\gamma}_\mA^{\rm II} =&~ -\hat{m}_\mA \left( 2  \phi_\mA\dfrac{\kappa_{\mA}}{\varepsilon_{\mA}}W'(\phi_\mA) -2 \phi_N\dfrac{\kappa_{N}}{\varepsilon_{N}}W'(\phi_N)\right.\nn\\
  &~\quad\quad\quad\left.-4 \kappa_{\mA}\varepsilon_{\mA}\phi_\mA\Delta \phi_\mA +4 \kappa_{N}\varepsilon_{N}\phi_N\Delta \phi_N \right.\nn\\
  &~\quad\quad\quad\left.+ 2\dfrac{\kappa_{\mA}}{\varepsilon_{\mA}}W(\phi_\mA)-2\dfrac{\kappa_{N}}{\varepsilon_{N}}W(\phi_N)\right.\nn\\
  &~\quad\quad\quad\left.-2 \kappa_{\mA}\varepsilon_{\mA}\|\nabla \phi_\mA\|^2+ 2\kappa_{N}\varepsilon_{N}\|\nabla \phi_N\|^2+ \left(\frac{1}{\rho_\mA}-\frac{1}{\rho_N}\right)p\right), 
\end{align}
for $\mA = 1, \dots, N-1$ and \eqref{eq: choice gamma N} for $\mA = N$.

Invoking relation \eqref{eq: chemical potential upsilon}, the corresponding volumetric free energy and associated chemical potential take the form:
\begin{subequations}\label{eq: selection GL form Psi II}
\begin{align}
    \hat{\Psi}_\mA^{\rm II} =&~2\dfrac{\rho_\mA\kappa_\mA}{\varepsilon_{\mA}}K(\phi_\mA) + 2\rho_\mA\kappa_\mA\varepsilon_\mA\phi_\mA\|\nabla \phi_\mA\|^2,\\
    K(\phi_\mA)=&~2\phi^3_\mA(1-\phi_\mA)^2,\\
    \mu^{\rm II}_\mA= &~  \phi_\mA  \rho_\mA \tau_\mA^{\rm II} +\rho_\mA \left(2\dfrac{\kappa_\mA}{\varepsilon_{\mA}}W(\phi_\mA) - 2\kappa_\mA\varepsilon_{\mA}\|\nabla \phi_\mA\|^2\right).
\end{align}
\end{subequations}

\begin{figure}[!ht]
\centering
\includegraphics[width=0.65\textwidth]{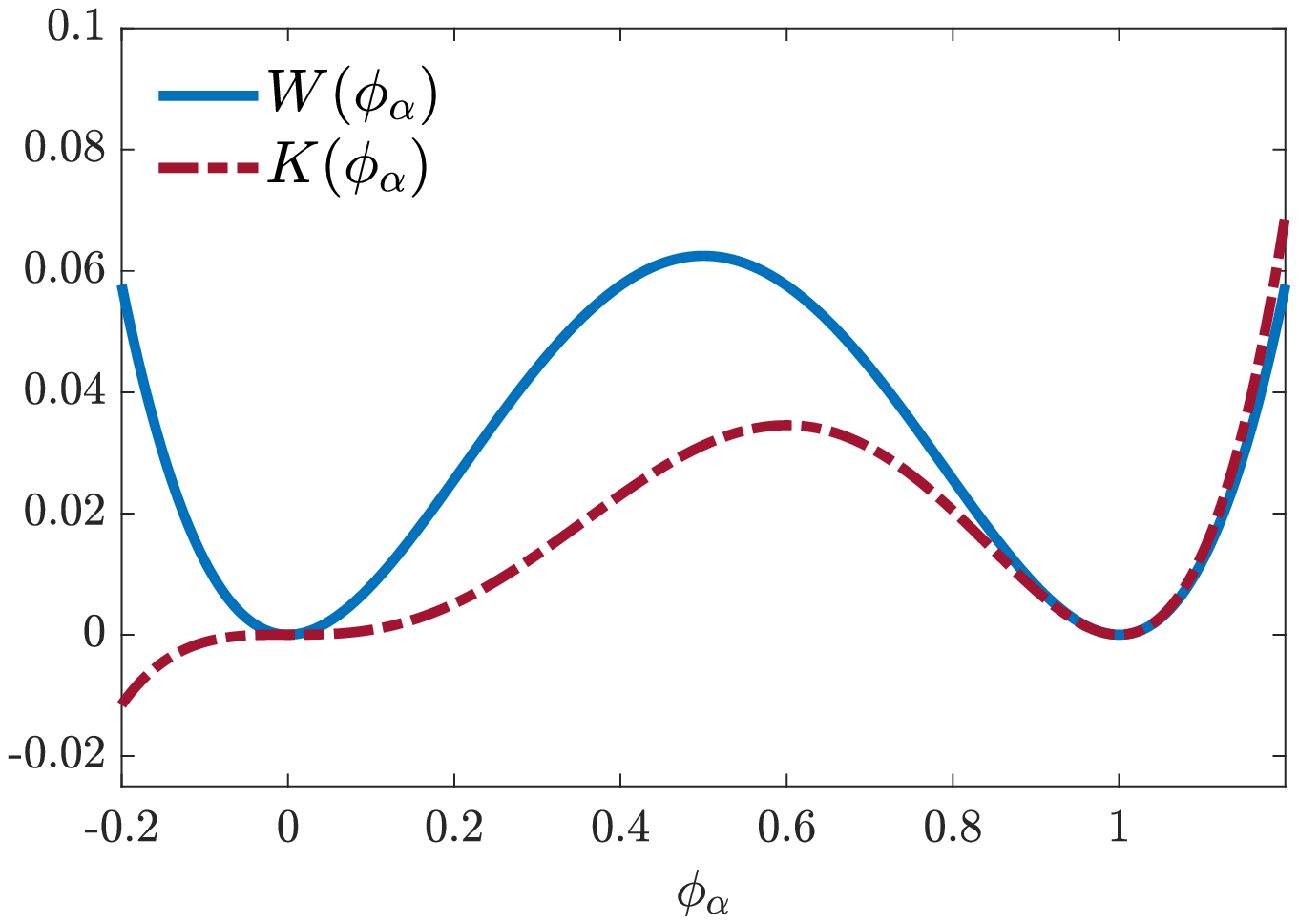}
\caption{The potentials $W=W(\phi_\mA)$ and $K=K(\phi_\mA)$.}
\label{fig: potentials}
\end{figure}

We visualize the potentials $W=W(\phi_\mA)$ and $K=K(\phi_\mA)$ in \cref{fig: potentials}. The potential $W=W(\phi_\mA)$ admits the well-known
symmetrical double-well shape, whereas $K=K(\phi_\mA)$ is a non-symmetric double-well.

\subsection{Dimensionless form}\label{subsec: dim less}
We perform non-dimensionalization based on the dimensionless variables:
  \begin{align}
    \bx^* :=&~ \frac{\bx}{L_0}, \quad \bv_\mA^* := \frac{\bv_\mA}{V_0}, \quad t^* := t \frac{V_0}{L_0}, \quad \tilde{\nu}_\mA^* := \frac{\tilde{\nu}_\mA}{\nu_\mA}, \quad p_\mA^* := \frac{p L_0}{a_\mA},  \nn\\
    \mu_\mA^* :=&~\frac{\mu_\mA L_0}{a_\mA},\quad   D^*_{\mA\mB}:=\dfrac{D_{\mA\mB}}{L_0 V_0},\quad \hat{m}_\mA^*:=\frac{a_\mA}{V_0\rho_\mA^2}\hat{m}_\mA,
  \end{align}
where $L_0, V_0, T_0$ and $\nu_\mA$ denote a characteristic length,
time, velocity, density, and constituent dynamic viscosity, respectively, and $a_\mA = \sigma_\mA$ and $a_\mA = \rho_\mA \kappa_\mA$ for models I and II respectively. The re-scaled system takes the form:
\begin{subequations}\label{eq: model dim less}
  \begin{align}
 \partial_{t^*} \phi_\mA + {\rm div}^*(\phi_\mA \bv_\mA^*) -\hat{\gamma}_\mA^* &=~ 0, \label{eq: model dim less: cont}\\
 \partial_{t^*} (\phi_\mA \bv_\mA^*) + {\rm div}^* \left( \phi_\mA \bv_\mA^*\otimes \bv_\mA^* \right)&\nn\\
 - \frac{1}{\mathbb{R}e_\mA}{\rm div}^* \left(\tilde{\nu}_\mA^* \left(2\mathbf{D}_\mA^* + \lambda_\mA {\rm div}^*\mathbf{v}_\mA^*\right)\right) &\nn\\
 + \dfrac{1}{\mathbb{W}{\rm e}_\mA}\phi_\mA \nabla^* \left(p^*_\mA+  \mu_\mA^*\right) 
 +\dfrac{1}{\mathbb{F}{\rm r}^2}\phi_\mA\boldsymbol{\jmath} &\nn\\  -\dfrac{1}{\mathbb{W}{\rm e}_\mA} p_\mA^*\displaystyle\sum_\mB \dfrac{\phi_\mA\phi_\mB}{D_{\mA\mB}^*}(\bv_\mB^*-\bv_\mA^*) +\boldsymbol{\beta}_\mA^*&=~ 0, \label{eq: model dim less: mom}
  \end{align}
\end{subequations}
for $\mA = 1,...,N$. 
Here $\nabla^*$, $\Delta^*$ and ${\rm div}^*$ denote the dimensionless spatial derivatives. The dimensionless variables are the constituent Reynolds number ($\mathbb{R}{\rm e}_\mA$), the Froude number ($\mathbb{F}{\rm r}$), the constituent Cahn number ($\mathbb{C}{\rm n}_\mA$) and the constituent Weber number ($\mathbb{W}{\rm e}_\mA$):
\begin{subequations}\label{eq: dimensionless quantities}
\begin{align}
     \mathbb{R}{\rm e}_\mA =&~ \frac{\rho_\mA V_0 L_0}{\nu_\mA},\\
     \mathbb{F}{\rm r} =&~ \frac{V_0}{\sqrt{b L_0}},\\
     \mathbb{C}{\rm n}_\mA =&~ \frac{\varepsilon_\mA}{L_0},\\
     \mathbb{W}{\rm e}_\mA =&~ \frac{\rho_\mA V_0^2L_0}{a_\mA}.
\end{align}
\end{subequations}
The dimensionless mass transfer terms read:
\begin{align}
  \hat{\gamma}_\mA^* =&~ -\hat{m}_\mA^* \left( \mu_\mA^*+p_\mA^*- \frac{\mathbb{W}{\rm e}_\mA}{\mathbb{W}{\rm e}_N}\left(\mu_N^*+p_N^*\right)\right),\quad \text{ for }\mA = 1, \dots, N-1,
\end{align}
and where
    \begin{align}
  \boldsymbol{\beta}_\mA^* =&~ \frac{1}{2} \hat{\gamma}_\mA^* \left(\mathbf{v}_\mA^* +\mathbf{v}_N^*\right),  \quad \text{ for }\mA = 1, \dots, N-1.
\end{align}
The free energies take the form:
\begin{subequations}\label{eq: dim less free energy}
  \begin{align}
  \hat{\Psi}^{I,*}_\mA =\hat{\psi}^{II,*}_\mA =&~  \dfrac{1}{\mathbb{C}{\rm n}_\mA\mathbb{W}{\rm e}_\mA}W(\phi_\mA) +  \dfrac{\mathbb{C}{\rm n}_\mA}{\mathbb{W}{\rm e}_\mA}\|\nabla \phi_\mA\|^2,\\
  \hat{\Psi}^{II,*}_\mA =&~  \dfrac{2}{\mathbb{C}{\rm n}_\mA\mathbb{W}{\rm e}_\mA}K(\phi_\mA) +  \dfrac{2\mathbb{C}{\rm n}_\mA}{\mathbb{W}{\rm e}_\mA}\phi_\mA\|\nabla \phi_\mA\|^2,
\end{align}
\end{subequations}
and the chemical potentials are: 
\begin{subequations}
    \begin{align}\label{eq: CP in form GL}
  \mu^{{\rm I},*}_\mA- \left( \dfrac{1}{\mathbb{C}{\rm n}_\mA}W'(\phi_\mA) -2\mathbb{C}{\rm n}_\mA\Delta^* \phi_\mA\right) &=~0,\\
  \mu^{{\rm II},*}_\mA- 2\phi_\mA\left( \dfrac{1}{\mathbb{C}{\rm n}_\mA}W'(\phi_\mA) -2\mathbb{C}{\rm n}_\mA\Delta^* \phi_\mA\right)&\nn\\
 -2 \left(\dfrac{1}{\mathbb{C}{\rm n}_\mA}W(\phi_\mA)- \mathbb{C}{\rm n}_\mA\|\nabla^* \phi_\mA\|^2\right) &=~0.
\end{align}
\end{subequations}
We suppress the star symbols in the remainder of this paper.

\subsection{Equilibrium profile}\label{subsec: equilibrium prof}
The static equilibrium profile of the model \eqref{eq: model dim less} is characterized by zero entropy production:
\begin{align}
 \displaystyle\sum_{\mA}  \mathscr{P}_\mA = 0.
\end{align}
From the equivalent form \eqref{eq: second law red 7} we find:
\begin{subequations}\label{eq: static eqs}
\begin{align}
     \tilde{\nu}_\mA \left( \mathbf{D} - \frac{1}{d} ({\rm div} \mathbf{v}_\mA) \mathbf{I}\right):\left(\mathbf{D} - \frac{1}{d} ({\rm div} \mathbf{v}_\mA) \mathbf{I}\right) =&~0, \label{eq: static eqs 1}\\
     \tilde{\nu}_\mA\left(\lambda_\mA + \frac{2}{d}\right)\left({\rm div} \mathbf{v}_\mA\right)^2 =&~0,\label{eq: static eqs 2}\\ 
     R_{\mA\mB}\|\bw_\mA-\bw_\mB\|^2 =&~0, \label{eq: static eqs 3}\\
     \hat{m}_\mA \left(g_\mA - g_N\right)^2 =&~0, \label{eq: static eqs 4}
\end{align}
\end{subequations}
for $\mA = 1, \dots, N$ in \eqref{eq: static eqs 1}-\eqref{eq: static eqs 2}, for $\mA, \mB = 1, \dots, N$ in \eqref{eq: static eqs 3}, and $\mA = 1, \dots, N-1$ in \eqref{eq: static eqs 4}.
Consider now the non-trivial case $0<\phi_\mA < 1$ and $\nu_\mA >0$. Since $\tilde{\nu}_\mA > 0$ we obtain from \eqref{eq: static eqs 1}-\eqref{eq: static eqs 1} that $\mathbf{v}_\mA = {\rm const}$ for all $\mA = 1, \dots N$. Next, since $R_{\mA\mB} \geq 0$ we get from \eqref{eq: static eqs 3} that $\mathbf{v}_\mA = \mathbf{v} = {\rm const}$ for all $\mA = 1, \dots N$. 
From \eqref{eq: static eqs 4} we obtain $g_1 = \dots, = g_N$ and $\hat{\gamma}_\mA = 0$ for all $\mA = 1, \dots, N$. As a consequence, from the mass balance equation \eqref{eq: model dim less: cont} we get $\grave{\phi}_\mA =0$. The viscous term and the last term in the momentum balance \eqref{eq: model dim less: mom} vanish due to $\mathbf{v}_\mA = {\rm const}$. Finally, the inertia terms in momentum balance \eqref{eq: model dim less: mom} vanish since:
\begin{align}
     \partial_{t} (\phi_\mA \bv_\mA) + {\rm div} \left( \phi_\mA \bv_\mA\otimes \bv_\mA \right) = \bv_\mA \grave{\phi}_\mA= 0.
\end{align}
The static equilibrium solution is now identified by the following relations:
\begin{subequations}\label{eq: eq relations}
\begin{align}
  \mu_\mA+p_\mA- \frac{\mathbb{W}{\rm e}_\mA}{\mathbb{W}{\rm e}_N}\left(\mu_N+p_N\right)&~=0,\quad \text{ for }\mA = 1, \dots, N-1,\label{eq: eq relations 1}\\
  \phi_\mA \nabla \left(p_\mA+  \mu_\mA +\dfrac{\mathbb{W}{\rm e}_\mA}{\mathbb{F}{\rm r}^2} y\right) &=~ 0,\quad \text{ for }\mA = 1, \dots, N.\label{eq: eq relations 2}
\end{align}    
\end{subequations}
\begin{remark}[Constituent body force]
  The equilibrium relations \eqref{eq: eq relations} are compatible due to the standing assumption of equal body forces ($\mathbf{b}_\mA=\mathbf{b}$ for $\mA = 1, \dots, N$).
\end{remark}
In scenario of a pure fluid ($\phi_\mA \equiv 1$), the thermodynamical pressure $\mu_\mA$ vanishes and we obtain $p_\mA = p_{\infty,\mA}- y\mathbb{W}{\rm e}_\mA/\mathbb{F}{\rm r}^2$, where $p_{\infty,\mA}$ is a constant equilibrium pressure. Consider now the non-trivial case ($0<\phi_\mA<1$) in absence of gravitational forces ($\mathbb{F}{\rm r}^{-2}=0$). 
The condition \eqref{eq: eq relations 1} implies that the quantity:
\begin{align}
\frac{1}{\mathbb{W}{\rm e}_\mA}\left( \mu_\mA+p_\mA\right) = C,
\end{align}
where $C$ is a constant independent of the constituent number. A solution is obtained by requiring $\mu_\mA = p_\mA = 0$. The zero pressure $p_\mA$ implies that momentum transfer is absent in equilibrium. The interface profiles $\phi_\mA = \phi_\mA^{\rm eq}(\xi)$ are determined by the differential equations:
\begin{subequations}\label{eq: Upsilon zero}
    \begin{align}
 0=&~ \mu_\mA^{\rm I} = \phi_\mA^{\rm eq} \left( \dfrac{1}{\mathbb{C}{\rm n}_\mA}W'(\phi_\mA^{\rm eq}) -2\mathbb{C}{\rm n}_\mA\Delta \phi_\mA^{\rm eq}\right),& \text{ for }\mA = 1, \dots, N\\
 0=&~ \mu_\mA^{\rm II} =2 \phi_\mA^{\rm eq} \left( \dfrac{1}{\mathbb{C}{\rm n}_\mA}W'(\phi_\mA^{\rm eq}) -2\mathbb{C}{\rm n}_\mA\Delta \phi_\mA^{\rm eq}\right)\nn\\
  &~+\dfrac{2}{\mathbb{C}{\rm n}_\mA}W(\phi_\mA^{\rm eq})- 2\mathbb{C}{\rm n}_\mA\|\nabla \phi_\mA^{\rm eq}\|^2,& \text{ for }\mA = 1, \dots, N, \\
  1=&~ \displaystyle\sum_\mA \phi_\mA.&
\end{align}
\end{subequations}

\begin{figure}[!ht]
\begin{subfigure}{0.49\textwidth}
\centering
\includegraphics[width=0.95\textwidth]{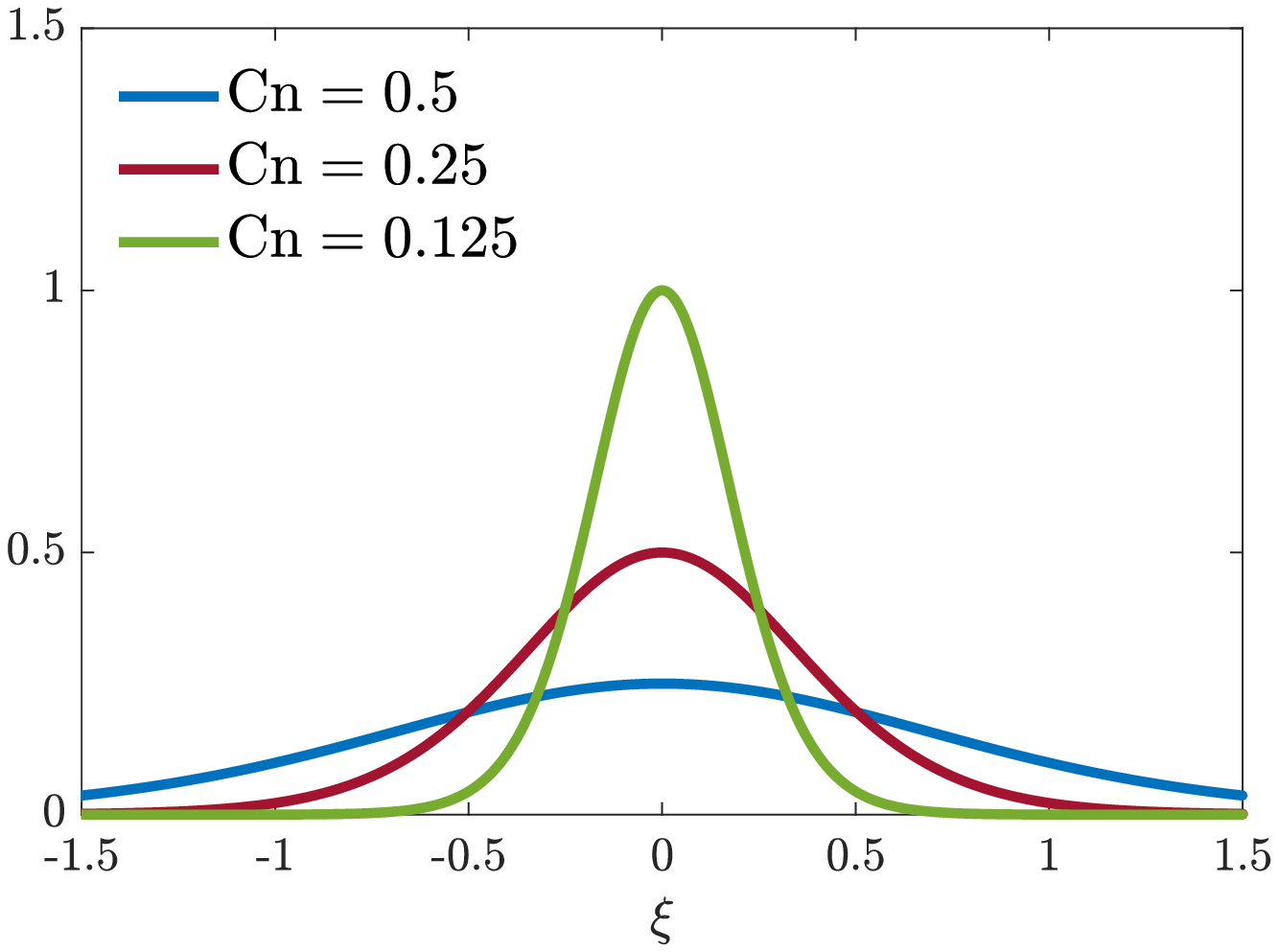}
\caption{$\mathbb{W}{\rm e}_\mA\hat{\Psi}_\mA^{\rm I}=\mathbb{W}{\rm e}_\mA\hat{\psi}_\mA^{\rm II}$.}
\end{subfigure}
\begin{subfigure}{0.49\textwidth}
\centering
\includegraphics[width=0.95\textwidth]{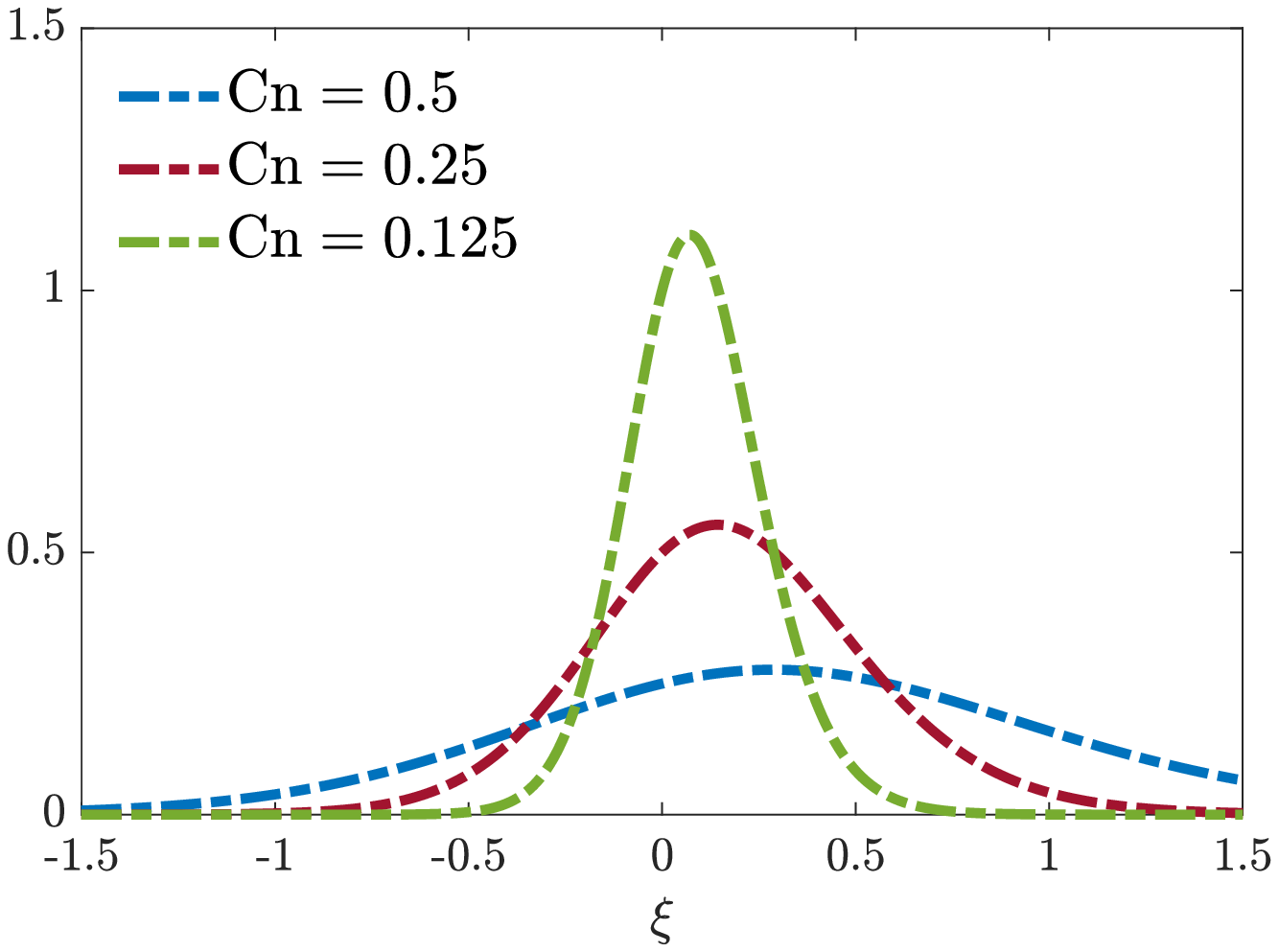}
\caption{$\mathbb{W}{\rm e}_\mA\hat{\Psi}_\mA^{\rm II}$.}
\end{subfigure}
\caption{The free energies for the equilibrium solution $\phi_\mA=\phi_\mA^{\rm eq}(\xi)$.}
\label{fig: free energies}
\end{figure}

We determine the explicit interface profiles in the one-dimensional situation. Denote with $\xi$ a spatial coordinate centered at the interface.
\begin{theorem}[Equilibrium profile]\label{thm: equilibrium prof}
    In absence of gravitational forces, the system \eqref{eq: model dim less} obeys in one-dimension the classical interface profile:
    \begin{align}\label{eq: interface prof}
  \phi_\mA = \phi_\mA^{\rm eq}(\xi) = \dfrac{1}{2}\left( 1 + \tanh\left(\dfrac{\pm \xi}{\mathbb{C}{\rm n}\sqrt{2}}\right)\right),
\end{align}
with $\mathbb{C}{\rm n}_{\mA}=\mathbb{C}{\rm n}$ for $\mA = 1, \dots, N$.
\end{theorem}
\begin{proof}
One may verify via substitution that the interface profile \eqref{eq: interface prof} satisfies the identities:
\begin{subequations}
  \begin{align}
       \dfrac{1}{\mathbb{C}{\rm n}_\mA}W'(\phi_\mA^{\rm eq}) -2\mathbb{C}{\rm n}_\mA \dfrac{{\rm d}^2 \phi_\mA^{\rm eq}}{{\rm d}\xi^2} = &~0,\\
      \dfrac{1}{\mathbb{C}{\rm n}_\mA}W(\phi_\mA^{\rm eq})- \mathbb{C}{\rm n}_\mA\left(\dfrac{{\rm d} \phi_\mA^{\rm eq}}{{\rm d}\xi}\right)^2 = &~0,
\end{align}
\end{subequations}
for $\mathbb{C}{\rm n}_{\mA}=\mathbb{C}{\rm n}, \mA = 1, \dots, N$.
\end{proof}
\cref{thm: equilibrium prof} conveys the shape of the interface profile, and moreover, it communicates that the interface width parameters need to be equal ($\mathbb{C}{\rm n}_{\mA}=\mathbb{C}{\rm n}, \mA = 1, \dots, N$). In the remainder of the paper we restrict to equal interface width parameters. As a consequence of the above identities we have
\begin{subequations}
  \begin{align}
    \hat{\Psi}_\mA^{\rm I}\left(\phi_\mA^{\rm eq}(\xi)\right) =\hat{\psi}^{\rm II}_\mA\left(\phi_\mA^{\rm eq}(\xi)\right)  =&~  \dfrac{2}{\mathbb{C}{\rm n}\mathbb{W}{\rm e}_\mA}W(\phi_\mA^{\rm eq})\nn\\
    =&~ \dfrac{1}{4\mathbb{C}{\rm n}\mathbb{W}{\rm e}_\mA}\left(1 - \tanh^2\left(\dfrac{\pm \xi}{\mathbb{C}{\rm n}\sqrt{2}}\right)\right)^2 ,\\
\hat{\Psi}_\mA^{\rm II}\left(\phi_\mA^{\rm eq}(\xi)\right) =&~ \dfrac{4}{\mathbb{C}{\rm n}\mathbb{W}{\rm e}_\mA}K(\phi_\mA^{\rm eq})\nn\\
=&~ \dfrac{1}{4\mathbb{C}{\rm n}\mathbb{W}{\rm e}_\mA}\left(1 + \tanh\left(\dfrac{\pm \xi}{\mathbb{C}{\rm n}\sqrt{2}}\right)\right)\times\nn\\
&~\left(1 - \tanh^2\left(\dfrac{\pm \xi}{\mathbb{C}{\rm n}\sqrt{2}}\right)\right)^2.
\end{align}
\end{subequations}
We visualize the free energies in \cref{fig: free energies}. The free energy of model I is symmetric around $0$, whereas the free energy of model I is non-symmetric. Both free energies collapse onto the interface for $\mathbb{C}{\rm n} \rightarrow 0$.

Finally, we introduce the (dimensionless) constituent surface tension coefficient as:
\begin{align}\label{eq: def theta}
 \hat{\Theta}_\mA= \displaystyle\int_{\mathbb{R}}  \hat{\Psi}_\mA\left(\phi_\mA^{\rm eq}(\xi)\right) {\rm d}\xi.
\end{align}
One may verify that the integral is the same for each of the two models:
  \begin{align}
    \displaystyle\int_{\mathbb{R}}  \hat{\Psi}_\mA^{\rm I}\left(\phi_\mA^{\rm eq}(\xi)\right) {\rm d}\xi =\displaystyle\int_{\mathbb{R}}  \hat{\Psi}_\mA^{\rm II}\left(\phi_\mA^{\rm eq}(\xi)\right) {\rm d}\xi = \dfrac{ \sqrt{2}}{3\mathbb{W}{\rm e}_\mA}.
\end{align}

\section{Connection with the Navier-Stokes Cahn-Hilliard model}\label{sec: connection}

In this section we explore the connection of the mixture model \eqref{eq: model FE} and the the Navier-Stokes Cahn-Hilliard model. We restrict ourselves to binary mixtures for the sake of clarity, and note that the extension to multi-constituent mixtures is straightforward. We discuss the connection for the diffuse-interface models outlined in \cref{sec: diffuse interface}. First, in \cref{subsec: NSCH} we lay down two particular forms of the NSCH model. Then, in \cref{subsec: connection} we analyze the connection of the components of the mixture model with the NSCH model. Finally, we discuss the connection of the complete models \cref{subsec: connection BL}.

\subsection{The Navier-Stokes Cahn-Hilliard model}\label{subsec: NSCH}

Restricting to two constituents, the volume fractions now constitute a single order parameter. We define this order parameter in the classical way as the difference of the volume fractions of the individual constituents: $\phi=\phi_1-\phi_2 \in [-1,1]$. Invoking \eqref{eq: def rho} and  \eqref{eq: sum phi} provides:
\begin{subequations}\label{eq: order parameter}
    \begin{align}
    \phi_1=&~\frac{1+\phi}{2}, \quad\quad \phi_2=\frac{1-\phi}{2},\\
    \rho(\phi) =&~ \frac{\rho_1(1+\phi)}{2}+\frac{\rho_2(1-\phi)
    }{2}.
\end{align}
\end{subequations}

We note that the NSCH model \eqref{eq: NSCH  model intro} is written is a form that directly allows the specification of a volume-measure-based Helmholtz free energy belonging to the constitutive class:
\begin{align}\label{eq: NSCH Psi}
  \bar{\Psi}=\bar{\Psi}(\phi,\nabla \phi).
\end{align}
On the other hand, it is also common to work with a Helmholtz free energy that is mass-measure-based:
\begin{align}\label{eq: NSCH psi}
  \bar{\psi}=\bar{\psi}(\phi,\nabla \phi).
\end{align}
We now present (equivalent) compact forms of the NSCH model, one suited for each of the two choices. 

To establish the connection between the two Helmholtz free energy classes we select the following natural identification:
\begin{align}
    \bar{\Psi}(\phi,\nabla \phi) \equiv \rho(\phi)\bar{\psi}(\phi,\nabla \phi).
\end{align}
Furthermore, we introduce chemical potentials associated with each of the constitutive classes:
\begin{subequations}\label{eq: chem pot I and II}
    \begin{align}
      \bar{\mu} =&~ \dfrac{\partial \bar{\Psi}}{\partial \phi}-{\rm div} \left(  \dfrac{\partial \bar{\Psi}}{\partial \nabla \phi} \right),\\
      \bar{\upsilon} =&~ \dfrac{\partial \bar{\psi}}{\partial \phi}-\frac{1}{\rho}{\rm div} \left(  \rho \dfrac{\partial \bar{\psi}}{\partial \nabla \phi} \right).
    \end{align}
\end{subequations}
With the aim of introducing the first compact form, we present a lemma analogous to \cref{lem: free energy contributions}.

\begin{lemma}[Compact form free energy contributions]\label{lem: free energy contributions mix}
The following identity holds:
\begin{align}\label{eq: ID free energy mix}
\phi \nabla \bar{\mu} =&~\nabla (\bar{\mu}\phi-\bar{\Psi}) + {\rm div} \left(\nabla \phi \otimes \dfrac{\partial \bar{\Psi}}{\partial \nabla\phi} \right).
\end{align}
\end{lemma}
\begin{proof}
  The proof is similar to that of \cref{lem: free energy contributions}.
\end{proof}
\begin{remark}
    The identity \eqref{eq: ID free energy mix} is often employed  in the particular scenario of the Ginzburg-Landau free energy. Here we note that it holds for the general constitutive class of the Helmholtz free energy.
\end{remark}

Applying \cref{lem: free energy contributions mix}, we arrive at the first form of the NSCH model:

\begin{subequations}\label{eq: NSCH  model sec5 form 1}
  \begin{align}
  \partial_t (\rho \bv) + {\rm div} \left( \rho \bv\otimes \bv \right) + \nabla p + \phi \nabla \bar{\mu} & \nn\\
    - {\rm div} \left(   \nu (2\mathbf{D}+\lambda({\rm div}\bv) \mathbf{I}) \right)-\rho\mathbf{b} &=~ 0, \label{eq: NSCH model v sec5}\\
 \partial_t \rho + {\rm div}(\rho \bv) &=~ 0, \label{eq: NSCH model rho sec5}\\
   \partial_t \phi + {\rm div}(\phi \bv) - {\rm div} \left(\bar{\mathbf{M}}\nabla (\bar{\mu}+\omega p)\right) +\zeta \bar{m} (\bar{\mu} + \omega p) &=~0,\label{eq: NSCH model phi sec5}
  \end{align}
\end{subequations}
Next, the second form of the NSCH model follows when switching to the mass-measure-based Helmholtz free energy in \eqref{eq: NSCH  model sec5 form 1}. To this purpose we introduce the relation between the chemical potentials \eqref{eq: chem pot I and II}.
\begin{lemma}[Relation chemical potentials]\label{lem: relation chem pot}
The chemical potentials \eqref{eq: chem pot I and II} are related as:
\begin{subequations}\label{eq: ID free energy mix2}
\begin{align}
\bar{\mu} = \rho \bar{\upsilon} + \bar{\psi} \frac{\rho_1-\rho_2}{2}.
\end{align}
\end{subequations}
\end{lemma}
\begin{proof}
  This follows from a straightforward substitution. For details we refer to \cite{eikelder2023unified}.
\end{proof}

Applying \cref{lem: relation chem pot}, we arrive at the second form of the NSCH model:
\begin{subequations}\label{eq: NSCH  model sec5 form 2}
  \begin{align}
  \partial_t (\rho \bv) + {\rm div} \left( \rho \bv\otimes \bv \right) + \nabla p + \phi \nabla \left(\rho \bar{\upsilon} + \bar{\psi} \frac{\rho_1-\rho_2}{2}\right) & \nn\\
    - {\rm div} \left(   \nu (2\mathbf{D}+\lambda({\rm div}\bv) \mathbf{I}) \right)-\rho\mathbf{b} &=~ 0, \label{eq: NSCH model v sec5 2}\\
 \partial_t \rho + {\rm div}(\rho \bv) &=~ 0, \label{eq: NSCH model rho sec5 2}\\
   \partial_t \phi + {\rm div}(\phi \bv) - {\rm div} \left(\bar{\mathbf{M}}\nabla \left(\rho \bar{\upsilon} + \bar{\psi} \frac{\rho_1-\rho_2}{2}+\omega p\right)\right) &\nn\\
   +\zeta m \left(\left(\rho \bar{\upsilon} + \bar{\psi} \frac{\rho_1-\rho_2}{2}\right) + \omega p\right) &=~0.\label{eq: NSCH model phi sec5 2}
  \end{align}
\end{subequations}

\begin{remark}[Variable transformation]
  One can apply a variable transformation in \eqref{eq: NSCH  model sec5 form 2} to absorb the term $\bar{\psi}(\rho_1-\rho_2)/2$ into the pressure $p$. For details we refer to \cite{eikelder2023unified}.
\end{remark}

Analogous to the diffuse-interface models in \cref{sec: diffuse interface}, we distinguish between a Ginzburg-Landau free energy that is either volume-measure-based, or mass-measure-based. It is our purpose to compare the associated models with the diffuse-interface models of \cref{sec: diffuse interface} (model I and model II). We also refer to the NSCH free energy models as model I and model II to emphasize this intend.\\

\noindent \textit{Model I}. The volume-measure-based Ginzburg-Landau free energy is given by:
\begin{subequations}\label{eq: selection GL form Psi NSCH}
    \begin{align}
    \bar{\Psi}^{\rm I} =&~ \dfrac{\sigma}{\varepsilon}F(\phi) + \dfrac{\sigma\varepsilon}{2}\|\nabla \phi\|^2,\\    
    F(\phi):=&~\frac{1}{4}(1-\phi^2)^2.\label{eq: def F}
\end{align}
\end{subequations}
where $F=F(\phi)$ represents a double-well potential, $\varepsilon$ is a (constant) interface thickness variable, and $\sigma$ is a (constant) variable related to the surface energy density. The chemical potential and mass transfer take the form:
\begin{subequations}\label{eq: CP + gamma in form I GL NSCH}
    \begin{align}
  \bar{\mu}^{\rm I} =&~   \dfrac{\sigma}{\varepsilon}F'(\phi) -\sigma\varepsilon\Delta \phi,\\
  \bar{\gamma}^{\rm I} =&~ - m \left(\bar{\mu}^{\rm I} + \omega p\right).
\end{align}
\end{subequations}

\noindent \textit{Model II}.
The mass-measured-based Ginzburg-Landau free energy reads:
\begin{align}\label{eq: selection GL form psi NSCH}
    \bar{\psi}^{\rm II} =&~ \dfrac{\kappa}{\varepsilon}F(\phi) + \dfrac{\kappa\varepsilon}{2}\|\nabla \phi\|^2,
\end{align}
where $F=F(\phi)$ is given in \eqref{eq: def F}. Also in this second model, the interface thickness variables $\varepsilon$ and surface energy density quantities $\kappa$ are assumed constant. The associated chemical potentials and mass transfer take the form:
\begin{subequations}\label{eq: CP + gamma in form I GL NSCH 2}
  \begin{align}
  \bar{\upsilon}^{\rm II}= &~  \bar{\tau}^{\rm II} -\dfrac{\kappa\varepsilon(\rho_1-\rho_2)}{2\rho} \|\nabla \phi \|^2,\\
  \bar{\tau}^{\rm II}:= &~  \dfrac{\kappa}{\varepsilon}F'(\phi) -\kappa\varepsilon\Delta \phi,\\
  \bar{\gamma}^{\rm II} =&~ -m \left( \rho \bar{\tau}^{\rm II}  + \dfrac{\rho_1-\rho_2}{2}\left(\dfrac{\kappa}{\varepsilon}F(\phi) - \dfrac{\kappa\varepsilon}{2}\|\nabla \phi\|^2\right)+ \omega p\right).
  \end{align} 
\end{subequations}

We now present the energy-dissipation property of the NSCH model. Introduce the global energy as the superposition of the Helmholtz free energy, kinetic energy and gravitational energy:
\begin{subequations}
    \begin{align}
      \bar{\mathscr{E}}(\Omega):= \displaystyle\int_\Omega \bar{\Psi} + \bar{\mathscr{K}} + \bar{\mathscr{G}}~{\rm d}\Omega, 
    \end{align}
\end{subequations}
where the Helmholtz free energy (\eqref{eq: NSCH Psi}) is specified in \eqref{eq: selection GL form Psi NSCH} and \eqref{eq: selection GL form psi NSCH}, the kinetic energy is given in \eqref{eq: kin avg}, and the gravitational energy is:
\begin{align}
  \bar{\mathscr{G}}: = \rho g y.
\end{align}
\begin{theorem}[Energy dissipation NSCH]\label{thm: NSCH energy balance}
Suppose that the NSCH model is equipped with the natural boundary conditions on $\Omega$:
\begin{subequations}
    \begin{align}
        \left(- p \mathbf{I} + \nu \left(2 \mathbf{D} + \lambda ({\rm div}\mathbf{v})\mathbf{I}\right)\right) \mathbf{n} =&~0,\\
        \nabla \phi \cdot \mathbf{n} =&~0,\\
        \left(\bar{\mathbf{M}} \nabla \left( \bar{\mu} + \omega p\right)\right)\mathbf{n} =&~0,
    \end{align}
\end{subequations}
where $\mathbf{n}$ denotes the outward unit normal, then the associated total energy satisfies the dissipation relation:
 \begin{align}
   \frac{{\rm d}}{{\rm d}t} \bar{\mathscr{E}}(\Omega)=&~- \displaystyle\int_\Omega \left( 2 \nu \left( \mathbf{D} - \frac{1}{d} ({\rm div} \mathbf{v}) \mathbf{I}\right):\left(\mathbf{D} - \frac{1}{d} ({\rm div} \mathbf{v}) \mathbf{I}\right)\right) ~{\rm d}\Omega\nn\\
    &~- \displaystyle\int \nu\left(\lambda + \frac{2}{d}\right)\left({\rm div} \mathbf{v}\right)^2 ~{\rm d}\Omega\nn\\
   &~ - \displaystyle\int_\Omega \nabla (\bar{\mu}+ \omega p)\cdot \left( \bar{\mathbf{M}}\nabla (\bar{\mu} + \omega p)\right)~{\rm d}\Omega\nn\\
   &~ - \displaystyle\int_\Omega \bar{m} \zeta (\bar{\mu}+ \omega p)^2~{\rm d}\Omega \leq 0.
 \end{align}
 \end{theorem}
The equilibrium profile of the model is characterized by zero energy evolution:
 \begin{align}
      \frac{{\rm d}}{{\rm d}t} \mathscr{E}(\Omega)= 0.
 \end{align}
Following a similar argumentation as in \cref{subsec: equilibrium prof}, in absence of gravitational forces one can deduce the equilibrium profile:
\begin{align}
    \phi = \phi^{\rm eq}(\xi) = \tanh\left(\dfrac{\pm \xi}{\varepsilon\sqrt{2}}\right),
\end{align}
where again $\xi$ is a coordinate centered at the interface ($\phi =0)$.

Lastly, consider the determination of the surface tension coefficient. Similar to \eqref{eq: def theta} we set:
\begin{subequations}
  \begin{align}
    \bar{\Theta}^{\rm I} =&~ \displaystyle\int_{\mathbb{R}} \bar{\Psi}^{\rm I}\left(\phi^{\rm eq}(\xi)\right) {\rm d}\xi,\\
    \bar{\Theta}^{\rm II} =&~ \displaystyle\int_{\mathbb{R}} \bar{\Psi}^{\rm II}\left(\phi^{\rm eq}(\xi)\right) {\rm d}\xi,
\end{align}
\end{subequations}
and note that the integrals are equal to:
\begin{subequations}
    \begin{align}
       \bar{\Theta}^{\rm I} =&~ \sigma \dfrac{2\sqrt{2}}{3},\\
       \bar{\Theta}^{\rm II} =&~ (\rho_1+\rho_2)\kappa \dfrac{\sqrt{2}}{3}.
    \end{align}
\end{subequations}
\subsection{Connection of the components of the mixture model }\label{subsec: connection}
To study the connection of the mixture model \eqref{eq: model FE} and the NSCH model \eqref{eq: NSCH  model sec5 form 1}, \eqref{eq: NSCH  model sec5 form 2}, it is useful to formulate the mixture model in terms of pure mixture quantities. The mixture quantities are the mixture velocity $\mathbf{v}$ (defined in \eqref{eq: mix velo}), the order parameter $\phi$ (defined \eqref{eq: order parameter}), and lastly a diffusive flux quantity defined as:
\begin{align}\label{eq: def J}
    \bJ :=&~ \trho_1 \bw_1-\trho_2\bw_2.
\end{align}
To formulate the mixture model \eqref{eq: model FE} in mixture quantities we introduce the variable transformations:
  \begin{subequations}\label{eq: var trans}
    \begin{align}
      \bv_1 =&~ \bv + \dfrac{\bJ}{2\trho_1}\\
      \bv_2 =&~ \bv - \dfrac{\bJ}{2\trho_2},
    \end{align}
  \end{subequations}
which follow from \eqref{eq: mix velo} and \eqref{eq: def J}.

In the remainder of this subsection we formulate the various energies and components of the mixture model \eqref{eq: model FE} in mixture quantities, and establish the connection with their counterparts in the NSCH model. We compare the quantities associated with the Ginzburg-Landau free energy model of \cref{subsec: GL Free Energy} with quantities of corresponding free energy model of \cref{subsec: NSCH}.\\

\noindent\underline{\textit{Kinetic energy}}. We recall from \eqref{eq: relation kin energies} that the kinetic energy of the mixture \eqref{eq: def sum K} may be decomposed as:
    \begin{align}\label{eq: relation kin energies 3}
      \mathscr{K} =&~ \bar{\mathscr{K}} + \displaystyle\sum_{\mA} \frac{1}{2} \trho_\mA \|\mathbf{w}_\mA\|^2.
\end{align}
The kinetic energy corresponding to the peculiar velocity is neglected in the NSCH model. The next lemma reformulates this kinetic energy in mixture quantities.
\begin{lemma}[Kinetic energy peculiar velocity]\label{lem: kin energy peculiar}
  The kinetic energy associated with the peculiar velocity takes the form:
    \begin{align}\label{eq: kin energy peculiar}
      \sum_{\mA=1,2}\tilde{\rho}_\mA\|\bw_\mA\|^2 = \dfrac{\rho\|\bJ\|^2}{2\rho_1 \rho_2(1-\phi^2)}.
  \end{align}
\end{lemma}
\begin{proof}
On the account of \eqref{eq: rel gross motion zero} we add a suitable partition of zero to the left-hand side and find:
    \begin{align}
      \sum_{\mA=1,2}\tilde{\rho}_\mA\|\bw_\mA\|^2 =&~ \bw_1\cdot \left( \trho_1\bw_1 + \trho_2\bw_2  \right) + \bw_2\cdot \left( \trho_1\bw_1 + \trho_2\bw_2  \right) \nn\\[-8pt]
      &~- \bw_1\cdot\trho_2\bw_2-\bw_2\cdot\trho_1\bw_1\nn\\
      =&~- \bw_1\cdot\trho_2\bw_2-\bw_2\cdot\trho_1\bw_1\nn\\
      =&~- \rho\bw_1\cdot\bw_2.
      \end{align}
Next, by recognizing the constituent diffusive flux we arrive at the result:
     \begin{align}
      \sum_{\mA=1,2}\tilde{\rho}_\mA\|\bw_\mA\|^2
      =&~ - \dfrac{\rho\bJ_1\cdot\bJ_2}{\trho_1 \trho_2}= \dfrac{\bJ \cdot \bJ}{4\trho_1 \trho_2}= \dfrac{\rho\|\bJ\|^2}{2\rho_1 \rho_2(1-\phi^2)}.
  \end{align}
\end{proof}

\noindent\underline{\textit{Gravitational energy}}. The gravitational energy of the mixture $\mathscr{G}$ coincides with the NSCH gravitational energy:
\begin{subequations}\label{eq: grav mix quantities}
\begin{align}
    \mathscr{G}_1 =&~ \rho_1 \frac{1+\phi}{2} b y,\\
    \mathscr{G}_2 =&~ \rho_2 \frac{1-\phi}{2} b y,\\
    \mathscr{G} =&~ \mathscr{G}_1 + \mathscr{G}_2 = \bar{\mathscr{G}}= \rho b y. \label{eq: G mix quantities}
\end{align}
\end{subequations}

\noindent\underline{\textit{Free energy}}. We define the mixture free energies as:
\begin{subequations}\label{eq: def mix free energy}
  \begin{align}
    \hat{\Psi}(\phi,\nabla \phi) =&~ \hat{\Psi}_1(\phi_1,\nabla \phi_1) + \hat{\Psi}_2(\phi_2,\nabla \phi_2),\\
    \rho \hat{\psi}(\phi,\nabla \phi) =&~ \trho_1 \hat{\psi}_1(\phi_1,\nabla \phi_1) + \trho_2 \hat{\psi}_2(\phi_2,\nabla \phi_2).
  \end{align}
\end{subequations}
We distinguish between the two models specified in \cref{subsec: GL Free Energy}.\\

\noindent \textit{Model I}. The constituent free energies \eqref{eq: selection GL form Psi} take the form:
\begin{subequations}\label{eq: selection GL form mix 1}
\begin{align}
    \hat{\Psi}_1^{\rm I} =&~ \dfrac{\sigma_{1}}{2\varepsilon}F(\phi) + \dfrac{\sigma_{1}\varepsilon}{4}\|\nabla \phi\|^2,\\
    \hat{\Psi}_2^{\rm I} =&~ \dfrac{\sigma_{2}}{2\varepsilon}F(\phi) + \dfrac{\sigma_{2}\varepsilon}{4}\|\nabla \phi\|^2,
\end{align}
\end{subequations}
where $F=F(\phi)$ is defined in \eqref{eq: def F}.
Inserting the Ginzburg Landau free energy \eqref{eq: selection GL form mix 1} into \eqref{eq: def mix free energy} we obtain:
\begin{align}\label{eq: selection GL form mix total}
    \hat{\Psi}^{\rm I} =&~ \left(\dfrac{\sigma_{1}}{2\varepsilon}+\dfrac{\sigma_{2}}{2\varepsilon}\right)F(\phi) + \dfrac{\sigma_{1}\varepsilon+\sigma_{2}\varepsilon}{4}\|\nabla \phi\|^2.
\end{align}
This form coincides with the standard Ginzburg Landau form \eqref{eq: selection GL form Psi NSCH} for the scenario $\sigma=\sigma_1=\sigma_2$:
\begin{align}\label{eq: Psi special choice}
    \hat{\Psi}^{\rm I} =&~ \bar{\Psi}^{\rm I} = \dfrac{\sigma}{\varepsilon}F(\phi) + \frac{\sigma\varepsilon}{2}\|\nabla \phi\|^2.
\end{align}
\noindent \textit{Model II}. The constituent free energies \eqref{eq: selection GL form psi} read:
\begin{subequations}\label{eq: selection GL form mix 2}
\begin{align}
    \hat{\psi}_1^{\rm II} =&~ \dfrac{\kappa_{1}}{\varepsilon}F(\phi) + \dfrac{\kappa_{1}\varepsilon}{2}\|\nabla \phi\|^2,\\
    \hat{\psi}_2^{\rm II} =&~ \dfrac{\kappa_{2}}{\varepsilon}F(\phi) + \dfrac{\kappa_{2}\varepsilon}{2}\|\nabla \phi\|^2.
\end{align}
\end{subequations}
Inserting the Ginzburg Landau free energy \eqref{eq: selection GL form mix 2} into \eqref{eq: def mix free energy} yields:
\begin{align}\label{eq: selection GL form mix total 2}
    \rho \hat{\psi}^{\rm II} =&~ \left(\dfrac{\rho_1\kappa_1}{2\varepsilon}+\dfrac{\rho_2\kappa_2}{2\varepsilon}\right)F(\phi) + \dfrac{\rho_1\kappa_1\varepsilon+\rho_2\kappa_2\varepsilon}{4}\|\nabla \phi\|^2 \nn\\
    &~+\left(\dfrac{\rho_1\kappa_1}{2\varepsilon}-\dfrac{\rho_2\kappa_2}{2\varepsilon}\right)\phi  F(\phi) + \dfrac{\rho_1\kappa_1\varepsilon-\rho_2\kappa_2\varepsilon}{4}\phi\|\nabla \phi\|^2.
\end{align}
In the special case $\kappa=\kappa_1=\kappa_2$ we retrieve the NSCH free energy:
\begin{align}
    \hat{\psi}^{\rm II} =&~ \bar{\psi}^{\rm II} = \dfrac{\kappa}{\varepsilon}F(\phi) + \frac{\kappa\varepsilon}{2}\|\nabla \phi\|^2.
\end{align}

\noindent\underline{\textit{Korteweg tensor}}. We differentiate between the two models specified in \cref{subsec: GL Free Energy}.\\

\noindent \textit{Model I}. The constituent Korteweg tensors read in mixture quantities:
\begin{subequations}\label{eq: const Kortweg Psi1}
  \begin{align}
    \nabla \phi_\mA \otimes \dfrac{\partial  \hat{\Psi}_1^{\rm I}}{\partial \nabla \phi_\mA}  =&~ \frac{\sigma_1\varepsilon}{2}\nabla \phi \otimes \nabla \phi,\\
    \nabla \phi_\mA \otimes \dfrac{\partial  \hat{\Psi}_2^{\rm I}}{\partial \nabla \phi_\mA} =&~ \frac{\sigma_2\varepsilon}{2}\nabla \phi \otimes \nabla \phi.
\end{align}
\end{subequations}
The superposition of the constituent Korteweg tensors yields:
\begin{align}
   \displaystyle\sum_{\mA=1,2} \nabla \phi_\mA \otimes \dfrac{\partial \hat{\Psi}_\mA^{\rm I}}{\partial \nabla \phi_\mA} = \nabla \phi \otimes \dfrac{\partial  \hat{\Psi}^{\rm I}}{\partial \nabla \phi} = \left(\frac{\sigma_1\varepsilon}{2}+\frac{\sigma_2\varepsilon}{2}\right)\nabla \phi \otimes \nabla \phi.
\end{align}
The first equality holds for all constituent classes $\hat{\Psi}^{\rm II}=\hat{\Psi}^{\rm I}(\phi,\nabla \phi)$, whereas the second follows from \eqref{eq: const Kortweg Psi1}. For the special case $\sigma=\sigma_1=\sigma_2$ we find the standard mixture Korteweg tensor:
\begin{align}
   \displaystyle\sum_{\mA=1,2} \nabla \phi_\mA \otimes \dfrac{\partial \hat{\Psi}_\mA^{\rm I}}{\partial \nabla \phi_\mA} = \sigma \varepsilon \nabla \phi \otimes \nabla \phi.
\end{align}

\noindent \textit{Model II}. The constituent Korteweg tensors read in mixture quantities:
\begin{subequations}\label{eq: const Kortweg Psi2}
  \begin{align}
    \nabla \phi_1 \otimes \dfrac{\partial  \hat{\psi}_1^{\rm II}}{\partial \nabla \phi_1}  =&~ \kappa_1\varepsilon\nabla \phi \otimes \nabla \phi,\\
    \nabla \phi_2 \otimes \dfrac{\partial  \hat{\psi}_2^{\rm II}}{\partial \nabla \phi_2} =&~ \kappa_2\varepsilon\nabla \phi \otimes \nabla \phi.
\end{align}
\end{subequations}
The superposition of the constituent Korteweg tensors yields:
\begin{align}
   \displaystyle\sum_{\mA=1,2} \nabla \phi_\mA \otimes \dfrac{\partial \hat{\Psi}_\mA^{\rm II}}{\partial \nabla \phi_\mA} &~= \nabla \phi \otimes \dfrac{\partial  \hat{\Psi}^{\rm II}}{\partial \nabla \phi} \nn\\
   &~= \left(\frac{\rho_1\kappa_1\varepsilon}{2}+\frac{\rho_2\kappa_2\varepsilon}{2}\right.\nn\\
   &\quad\quad\left.+\phi\frac{\rho_1\kappa_1\varepsilon}{2}-\phi\frac{\rho_2\kappa_2\varepsilon}{2}\right)\nabla \phi \otimes \nabla \phi.
\end{align}
In the scenario $\kappa=\kappa_1=\kappa_2$ the mixture Korteweg tensor reduces to:
\begin{align}
   \displaystyle\sum_{\mA=1,2} \nabla \phi_\mA \otimes \dfrac{\partial \hat{\Psi}_\mA^{\rm I}}{\partial \nabla \phi_\mA} = \rho \kappa \varepsilon \nabla \phi \otimes \nabla \phi.
\end{align}

\noindent\underline{\textit{Chemical potential}}. Likewise the other terms involving the free energy, we separate the two modeling choices specified in \cref{subsec: GL Free Energy}.\\

\noindent \textit{Model I}. The chemical potentials take the form:
\begin{subequations}\label{eq: chem mix I}
    \begin{align}
\mu_1^{\rm I} = &~\dfrac{\sigma_{1}}{\varepsilon}F'(\phi) -\sigma_{1}\varepsilon\Delta \phi\\
\mu_2^{\rm I} = &~-\dfrac{\sigma_{2}}{\varepsilon}F'(\phi) +\sigma_{2}\varepsilon\Delta \phi.
\end{align}
\end{subequations}
In the case $\sigma=\sigma_1=\sigma_2$ we arrive at:
    \begin{align}
\mu_1^{\rm I} = -\mu_2^{\rm I} = \bar{\mu}^{\rm I} = \dfrac{\sigma}{\varepsilon}F'(\phi) -\sigma\varepsilon\Delta \phi.
\end{align}
\noindent \textit{Model II}. The associated chemical potentials take the form:
\begin{subequations}\label{eq: chem mix II}
    \begin{align}
\mu_1^{\rm II} = &~\dfrac{1+\phi}{2}\rho_1\tau_1 +\rho_1\left(\dfrac{\kappa_1}{\varepsilon}F(\phi) -\frac{\kappa_1\varepsilon}{2}\|\nabla \phi\|^2\right),\\
\mu_2^{\rm II} = &~\dfrac{1-\phi}{2}\rho_2\tau_2+\rho_2\left(\dfrac{\kappa_2}{\varepsilon}F(\phi) -\frac{\kappa_2\varepsilon}{2}\|\nabla \phi\|^2\right),\\
\tau_1^{\rm II} = &~\dfrac{2\kappa_1}{\varepsilon}F'(\phi) -2\kappa_1\varepsilon\Delta \phi,\\
\tau_2^{\rm II} = &~-\dfrac{2\kappa_2}{\varepsilon}F'(\phi) +2\kappa_2\varepsilon\Delta \phi,
\end{align}
\end{subequations}
In the case $\kappa=\kappa_1=\kappa_2$ we arrive at:
\begin{subequations}
    \begin{align}
  \mu_1^{\rm II} =&~ \rho_1 (1+\phi) \bar{\tau}^{\rm II} + \rho_1\left(\dfrac{\kappa}{\varepsilon}F(\phi) -\frac{\kappa\varepsilon}{2}\|\nabla \phi\|^2\right),\\
  \mu_2^{\rm II} =&~ - \rho_2 (1-\phi) \bar{\tau}^{\rm II} + \rho_2\left(\dfrac{\kappa}{\varepsilon}F(\phi) -\frac{\kappa\varepsilon}{2}\|\nabla \phi\|^2\right).
\end{align}
\end{subequations}
The free energy contributions take the form:
\begin{align}
     \displaystyle\sum_{\mA=1,2} \phi_\mA \nabla \mu_\mA^{\rm I} =&~ \frac{\phi}{2} \nabla  \left(\mu^{\rm I}_1-\mu^{\rm I}_2\right) + \frac{1}{2}\nabla \left(\mu^{\rm I}_1+\mu^{\rm I}_2\right).
\end{align}
\begin{lemma}[Reduction free energy contribution]\label{lem: reduction free energy contribution}
  In case of equal parameters $\sigma=\sigma_1=\sigma_2$ (model I), and $\kappa = \kappa_1 = \kappa_2$ (model II), the surface tension contributions reduce to: 
  \begin{subequations} 
  \begin{align}
     \displaystyle\sum_{\mA=1,2} \phi_\mA \nabla \mu_\mA^{\rm I} =&~ \phi \nabla \bar{\mu}^{\rm I},\\
     \displaystyle\sum_{\mA=1,2} \phi_\mA \nabla \mu_\mA^{\rm II} =&~ \phi \nabla \left( \rho \bar{\upsilon}^{\rm II}  + \bar{\psi}^{\rm II} \frac{\rho_1-\rho_2}{2} \right) + \mathbf{c},\label{eq: sur ten 2}\\
     \mathbf{c} =&~ \nabla \left((\trho_1-\trho_2)\bar{\tau}^{\rm II}+\frac{\rho_1+\rho_2}{2}\left(\frac{\kappa}{\varepsilon} F(\phi) -\dfrac{\kappa\varepsilon}{2}\|\nabla \phi\|^2\right)\right).
  \end{align}
  \end{subequations}
\end{lemma}
\begin{proof}
  This is a straightforward consequence of the variable transformation \eqref{eq: order parameter} and the form of the chemical potentials \eqref{eq: chem mix I} and \eqref{eq: chem mix II}.
\end{proof}
\cref{lem: reduction free energy contribution} conveys that for free energy model I the surface tension contribution coincides with that of the NSCH model. On the other hand, for model II it does not match with the NSCH model due to the presence of $\mathbf{c}$ in \eqref{eq: sur ten 2} (which is in general not zero).\\


\noindent\underline{\textit{Mass transfer}}. On the account of the balance \eqref{eq: balance mass fluxes}, we introduce a single mass transfer quantity $\hat{\gamma}$ that is related to the constituent mass transfer quantities via:
\begin{align}
    \hat{\gamma} = \hat{\gamma}_1 - \hat{\gamma}_2, \quad\quad \hat{\gamma}_1 = \frac{1}{2}\hat{\gamma}, \quad\quad \hat{\gamma}_2= -\frac{1}{2}\hat{\gamma}.   
\end{align}
We distinguish the two free energy models specified in \cref{subsec: GL Free Energy}.\\

\noindent \textit{Model I}. Substitution of the order parameter into \eqref{eq: gamma GL form I} provides:
\begin{align}\label{eq: gamma GL form I mix}
\hat{\gamma}^{\rm I} =&~ -\hat{m} \left( \left(\dfrac{\sigma_{1}}{\rho_1\varepsilon}+\dfrac{\sigma_{2}}{\rho_2\varepsilon}\right)F'(\phi) -\left(\dfrac{\sigma_{1}\varepsilon}{\rho_1}+\dfrac{\sigma_{2}\varepsilon}{\rho_2}\right)\Delta \phi\right.\nn\\
  &~\quad\quad\quad\quad\left.+\left(\dfrac{1}{\rho_1}-\dfrac{1}{\rho_2}\right)p\right),
\end{align}
where $\hat{m}=2\hat{m}_1=2\hat{m}_2$. In the scenario $\sigma=\sigma_1=\sigma_2$ the mass transfer reduces to the NSCH mass transfer:
\begin{align}\label{eq: gamma GL form I mix equal}
\hat{\gamma}^{\rm I} =&~ \bar{\gamma}^{\rm I} = -\bar{m} \left( \bar{\mu}^{\rm I}+\omega p\right),
\end{align}
with $\bar{m} = \hat{m} (\rho_1^{-1}+\rho_2^{-1})$.\\

\noindent \textit{Model II}. Substitution of the order parameter into \eqref{eq: gamma GL form II} provides:
\begin{align}\label{eq: gamma GL form II mix}
\hat{\gamma}^{\rm II} =&~ -\hat{m} \left(  \left(\dfrac{\kappa_{1}}{\varepsilon}+\dfrac{\kappa_{2}}{\varepsilon}\right)F'(\phi) +  \left(\dfrac{\kappa_{1}}{\varepsilon}- \dfrac{\kappa_{2}}{\varepsilon}\right)\phi F'(\phi)\right.\nn\\
  &~\quad\quad\quad\left.- \left(\kappa_{1}\varepsilon + \kappa_2\varepsilon\right)\Delta \phi -\left(\kappa_{1}\varepsilon - \kappa_2\varepsilon\right)\phi\Delta \phi \right.\nn\\
  &~\quad\quad\quad\left.+ \left(\dfrac{\kappa_{1}}{\varepsilon}-\dfrac{\kappa_{2}}{\varepsilon}\right)F(\phi)- \left(\dfrac{\kappa_{1}\varepsilon}{2}-\dfrac{\kappa_{2}\varepsilon}{2}\right)\|\nabla \phi\|^2\right.\nn\\
  &~\quad\quad\quad\left. + \left(\frac{1}{\rho_1}-\frac{1}{\rho_2}\right)p\right),
\end{align}
where $\hat{m}=2\hat{m}_1=2\hat{m}_2$. In the scenario $\kappa=\kappa_1=\kappa_2$ the mass flux reduces to:
\begin{align}\label{eq: gamma GL form II mix equal}
\hat{\gamma}^{\rm II} =&~ -\breve{m}\left( \frac{2\rho_1\rho_2}{\rho_1+\rho_2}\bar{\tau}^{\rm II}+ \omega p\right),
\end{align}
with $\breve{m}=m(\rho_1+\rho_2)/(\rho_1\rho_2)$. This does in general not match with the NSCH mass transfer. However, in the density matching case $\rho_1=\rho_2=\rho$ it reduces to the NSCH mass transfer $\hat{\gamma}^{\rm II}=\bar{\gamma}^{\rm II}$.\\ 

\noindent\underline{\textit{Momentum transfer}}. Based on the balance \eqref{eq: balance momentum fluxes}, we introduce the momentum transfer $\hat{\gamma}$ related to the constituent momentum transfer quantities via:
\begin{align}
    \hat{\bpi}=\hat{\bpi}_1-\hat{\bpi}_2, \quad\quad \hat{\bpi}_1 = \frac{1}{2}\hat{\bpi}, \quad\quad \hat{\bpi}_2= -\frac{1}{2}\hat{\bpi}.    
\end{align}
Inserting the order parameter and denoting $D = D_{12} =  D_{21}$, we obtain:
\begin{align}
  \bpi =&~ p \nabla \phi-\frac{\rho p}{2D\rho_1\rho_2}\bJ + \frac{1}{2} \hat{\gamma} \mathbf{v}  + \frac{\hat{\gamma}}{2}\left(\frac{1}{\rho_1(1+\phi)}-\frac{1}{\rho_2(1-\phi)}\right)\mathbf{J},
\end{align}
where the last member vanishes when $\phi = \pm 1$.\\

\noindent\underline{\textit{Viscous stress tensor}}. 
Invoking the variable transformation \eqref{eq: var trans}, the superposition of the viscous components of the stress tensors admits the form:
\begin{subequations}\label{eq: mix viscosity tensor}
    \begin{align}
  \sum_{\mA=1,2} \tilde{\nu}_\mA \left(2\mathbf{D}_\mA + \lambda_\mA ({\rm div}\mathbf{v}_\mA) \mathbf{I}\right) = &~ \nu \left(2\mathbf{D}+\lambda {\rm div}\bv \right) \nn\\[-8pt]
  &~+ \hat{\nu} \left( 2\mathbf{A} + \lambda \left({\rm div}\bJ\right) \mathbf{I}\right)\nn\\
  &~+ \breve{\nu} \left( 2\mathbf{B}+\lambda\left(\bJ\cdot \nabla \phi\right)\mathbf{I}\right),  
\end{align}
\end{subequations}
where we have introduced the viscosity quantities:
\begin{subequations}
  \begin{align}
      \nu =&~ \nu_1\frac{1+\phi}{2} + \nu_2\frac{1-\phi}{2},\\
      \hat{\nu} =&~  \dfrac{\nu_1}{2\rho_1} -\dfrac{\nu_2}{2\rho_2} ,\\
      \breve{\nu} =&~- \dfrac{\nu_1}{2\rho_1(1+\phi)} +  \dfrac{\nu_2}{2\rho_2(1-\phi)},
  \end{align}
\end{subequations}
the symmetric tensors:
\begin{subequations}
  \begin{align}
      \mathbf{D} =&~ \frac{1}{2}\left(\nabla \bv + (\nabla \bv)^T\right),\\
      \mathbf{A} =&~ \frac{1}{2}\left(\nabla \bJ + (\nabla \bJ)^T\right),\\
      \mathbf{B} =&~ \frac{1}{2}\left(\bJ\otimes \nabla \phi + \nabla \phi\otimes \bJ\right),
  \end{align}
\end{subequations}
and we have set $\lambda = \lambda_1=\lambda_2$. In establishing the above form we have made use of the identities:
\begin{subequations}
  \begin{align}
      \nabla \bv_1 =&~ \nabla \bv + \dfrac{1}{\rho_1(1+\phi)}\nabla \bJ - \dfrac{1}{\rho_1(1+\phi)^2} \mathbf{J} \otimes \nabla \phi, \\
      \nabla \bv_2 =&~\nabla \bv - \dfrac{1}{\rho_2(1-\phi)}\nabla \bJ + \dfrac{1}{\rho_2(1-\phi)^2} \mathbf{J} \otimes \nabla \phi.
  \end{align}
\end{subequations}
Each of the three members of the viscous stress tensor \eqref{eq: mix viscosity tensor} appears in the classical form of a symmetric tensor and $\lambda \mathbf{I}$ times its trace. The form \eqref{eq: mix viscosity tensor} conveys that the mixture viscous stress term is composed of contribution solely associated with the mixture velocity $\bv$, and a part in terms of the diffusive velocity $\bJ$. The first contribution is precisely the viscous stress tensor in the Navier-Stokes Cahn-Hilliard model. In contrast, the second contribution represents diffusion with respect to the peculiar velocity. This contribution is absent in the Navier-Stokes Cahn-Hilliard model.\\

\noindent\underline{\textit{Peculiar velocity stress component}}. With the aim of expressing the peculiar velocity component of the stress in mixture variables, we introduce the following lemma.

\begin{lemma}[Symmetry dyadic product peculiar velocity]\label{lem: sym dyadic}
The peculiar velocity dyadic product is symmetric: 
\begin{align}
    \bw_1 \otimes \bw_2 =  \bw_2 \otimes \bw_1.
\end{align}
\end{lemma}
\begin{proof}
This follows from the sequences of identities:
    \begin{align}
      \bw_1 \otimes \bw_2 = &~ \left( \bv_1-\bv\right) \otimes \left( \bv_2-\bv\right)\nn\\
      = &~ \bv_1\otimes\bv_2 - \frac{1}{\rho} \bv_1\otimes\left(\trho_1\bv_1+\trho_2\bv_2\right) -  \frac{1}{\rho}\left(\trho_1\bv_1+\trho_2\bv_2\right)\otimes\bv_2 +  \bv\otimes\bv\nn\\
      = &~  - \frac{\trho_1}{\rho} \bv_1\otimes\bv_1 -  \frac{\trho_2}{\rho}\bv_2\otimes\bv_2 +  \bv\otimes\bv.
  \end{align}
\end{proof}

We may now write the peculiar velocity component in mixture quantities. 
\begin{lemma}[Peculiar velocity component stress]\label{lem: stress peculiar}
  The peculiar velocity component of the stress takes the form:
    \begin{align}\label{eq: pec vel stress}
      \sum_{\mA=1,2}\tilde{\rho}_\mA\bw_\mA\otimes\bw_\mA = \dfrac{\rho\bJ\otimes \bJ}{2\rho_1 \rho_2(1-\phi^2)}.
  \end{align}
\end{lemma}
\begin{proof}
The proof goes similar as that of \cref{lem: kin energy peculiar} and relies on \cref{lem: sym dyadic}.

\end{proof}
This contribution represents the inertia of the diffusive flux. It is not present in the NSCH model.

\subsection{Connection of the complete models}\label{subsec: connection BL}

We start with the mass balance laws. The mixture mass balance law
\begin{align}\label{eq: mix mass balance}
    \partial_t \rho + {\rm div}\left( \rho \mathbf{v} \right) = 0,
\end{align}
as presented in \eqref{eq: local mass balance mix}, is identical in the mixture model \eqref{eq: model FE} and the NSCH models \eqref{eq: NSCH  model sec5 form 1} and \eqref{eq: NSCH  model sec5 form 2}. Next, the phase equation 
formulated in mixture quantities
follows from \eqref{eq: model FE: cont}:
\begin{align}\label{eq: phase eq mix model}
  \partial_t \phi + {\rm div}(\phi \mathbf{v}) + {\rm div} \mathbf{h} - \zeta \gamma = 0,
\end{align}
where we have introduced the diffusive flux quantity:
\begin{align}\label{eq: def h}
    \bh = \phi_1\bw_1 - \phi_2 \bw_2.
\end{align}
This equation is \textit{not} of Cahn-Hilliard type. The phase equation \eqref{eq: phase eq mix model} does not contain a chemical potential or pressure variable. This sets it apart from it NSCH counterpart in which the diffusive flux $\mathbf{h}$ is replaced by the constitutive model:
\begin{subequations}\label{eq: model h}
\begin{align}
  \bar{\bh}^{\rm I} = &~ - \bar{\mathbf{M}}\nabla (\bar{\mu} + \omega p),\quad\quad\quad\quad&(\text{Model I})\\
  \bar{\bh}^{\rm II} =&~ - \bar{\mathbf{M}}\nabla \left(\rho \bar{\upsilon} + \bar{\psi} \frac{\rho_1-\rho_2}{2}+\omega p\right).\quad\quad\quad\quad&(\text{Model II})
\end{align}
\end{subequations}
The diffusive flux \eqref{eq: def h} and the constitutive model \eqref{eq: model h} both vanish in equilibrium. 
On the other hand, the mass transfer term of the mixture model and the NSCH model is of similar type. In the scenario of model I with equal modeling parameters ($\sigma_1=\sigma_2$) it coincides with the NSCH mass transfer (see \cref{subsec: connection}).

\begin{remark}[Diffusive fluxes] 
The diffusive fluxes $\bJ$ and $\bh$ constitute a single unknown in the system, since they are related as $\bJ = 2\rho_1\rho_2\bh/(\rho_1+\rho_2)$. For a proof we refer to \cite{eikelder2023unified}.   
\end{remark}

Next, we focus on the mixture momentum equation which follows from the superposition of the constituent momentum balance equations \eqref{eq: model FE: mom}:
\begin{align}\label{eq: mom mix}
 \partial_t \mathbf{m} + {\rm div} \left( \mathbf{m}\otimes \bv \right) + \nabla p 
  - {\rm div}\left(\nu \left(2\mathbf{D}+\lambda {\rm div}\bv \right)\right)-\rho\mathbf{b}  &\nn\\
  + \frac{\phi}{2} \nabla  \left(\mu^{\rm I}_1-\mu^{\rm I}_2\right) + \frac{1}{2}\nabla \left(\mu^{\rm I}_1+\mu^{\rm I}_2\right)&\nn\\
 - {\rm div}\left( \hat{\nu} \left( 2\mathbf{A} + \lambda \left({\rm div}\bJ\right) \mathbf{I}\right)+ \breve{\nu} \left( 2\mathbf{B}+\lambda\left(\bJ\cdot \nabla \phi\right)\mathbf{I}\right)\right) &\nn\\
  + {\rm div} \left( \dfrac{\rho\bJ\otimes \bJ}{2\rho_1 \rho_2(1-\phi^2)} \right)
    &=~ 0. 
\end{align}
where we have substituted the expressions for viscous, and peculiar velocity contributions. The first line matches with the NSCH model. The second line consists of free energy terms. In case of equal modeling parameters, it reduces for model I to the free energy contribution in the NSCH model. This does not apply to the second model. The members of the last two lines are absent in the NSCH linear momentum equation. These terms are all linked to the diffusive flux. The diffusive flux in the mixture model is described by an evolution, whereas in the NSCH model it is determined by the constitutive model \eqref{eq: model h}. This is related to the usage of the energy-dissipation statement modeling restriction of the NSCH model, instead of the second law of thermodynamics adopted for the mixture model. It precludes the need of a constitutive model for the momentum transfer. The system described by the mixture mass balance \eqref{eq: mix mass balance}, the phase equation \eqref{eq: phase eq mix model}, the linear momentum equation \eqref{eq: mom mix}, augmented with the evolution equation of the diffusive flux (see \cite{eikelder2023unified}) is equivalent to the mixture model \eqref{eq: model FE} (for the diffuse-interface models of \cref{sec: diffuse interface}).

The mixture model and the NSCH model share the same one-dimensional equilibrium profile:
\begin{align}
    \phi = \phi^{\rm eq}(\xi) = \tanh\left(\dfrac{\pm \xi}{\varepsilon\sqrt{2}}\right).
\end{align} 
We consider the surface tension coefficient and define for both models:
  \begin{align}
    \hat{\Theta} := \hat{\Theta}_1 + \hat{\Theta}_2.
\end{align}
This results in:
\begin{subequations}
    \begin{align}
       \hat{\Theta}^{\rm I} =&~ (\sigma_1+\sigma_2) \dfrac{\sqrt{2}}{3},\\
       \hat{\Theta}^{\rm II} =&~ (\rho_1\kappa_1+\rho_2\kappa_2)\dfrac{\sqrt{2}}{3}.
    \end{align}
\end{subequations}
For equal parameters $\sigma_1=\sigma_2=\sigma$ and $\kappa_1 =\kappa_2 =\kappa$ these integrals match with the NSCH surface tension coefficients:
\begin{subequations}
    \begin{align}
       \hat{\Theta}^{\rm I} =&~ \bar{\Theta}^{\rm I} = \sigma \dfrac{2\sqrt{2}}{3},\\
       \hat{\Theta}^{\rm II} =&~ \bar{\Theta}^{\rm II} =  (\rho_1+\rho_2)\kappa \dfrac{\sqrt{2}}{3}.
    \end{align}
\end{subequations}

Lastly, we summarize the comparison of the mixture model and the NSCH model in \cref{table: overview comparison models}.
\begin{center}
\begin{table}[h!]
\centering
{\small
\begin{tabular}{c c c}
                      & \textbf{Mixture model} & \textbf{NSCH model}  \\[6pt] \thickhline\\[-4pt]

Mixture theory            & {\color{darkgreen}\cmark}         & {\color{red}\xmark}   \\[6pt] \hline 
Modeling restriction & Second law  & Energy-dissipative  \\[6pt] \hline
\# mass balance laws & $N$  & $N$ \\[6pt] \hline
\# momentum balance laws & $N$  & $1$ \\[6pt] \hline
Diffusive flux & Evolution equation  & Constitutive model \\[6pt] \hline
Interface profile& Tangent hyperbolic & Tangent hyperbolic  \\[6pt] \hline
\end{tabular}}
\caption{Comparison mixture model and NSCH model for $N$ constituents. With the term `mixture theory' we indicate whether the model is compatible with mixture theory. Next, energy-dissipative refers to the energy-dissipative property of NSCH model. Finally, in the last line we note that both models admit the standard tangent hyperbolic interface profile for the Ginzburg-Landau free energy.}
\label{table: overview comparison models}
\end{table}
\end{center}

\section{Conclusion}\label{sec: discussion} 

In this paper, we presented a thermodynamical consistent diffuse-interface incompressible mixture model. Starting from the continuum theory of mixtures we derived a constitutive modeling restriction that is compatible with the second law of thermodynamics. Subsequently, we selected constitutive models that satisfy this modeling restriction. To close the mixture model, we presented two diffuse-interface models, each associated with a particular Helmholtz free energy. Finally, we studied in detail the connection with the Navier-Stokes Cahn-Hilliard model (see \cref{table: overview comparison models} for an overview).

While the diffuse-interface mixture models we have set out are helpful in the study of evolution of incompressible mixtures, we certainly do not claim that these are sufficient. We outline two main avenues of potential future research. The first avenue is the rigorous mathematical analysis of the models, and the study of the sharp interface asymptotics. This sharp interface analysis is of different type than of the Navier-Stokes Cahn-Hilliard model. Indeed, the proposed mixture models are not of Cahn-Hilliard type and do not contain a mobility parameter. Furthermore, to assess the behavior of solutions of the mixture model, it is essential to develop suitable numerical algorithms. In particular, it is worthwhile to compare numerical solutions of the mixture model with those of the Navier-Stokes Cahn-Hilliard model.

\section*{Acknowledgments}
MtE acknowledges support from the German Research Foundation (Deutsche Forschungsgemeinschaft DFG) via the Walter Benjamin project EI 1210/1-1. 
The research by KvdZ was supported by the Engineering and Physical Sciences Research Council (EPSRC), UK, under Grants EP/T005157/1 and EP/W010011/1. DS gratefully acknowledges support from the German Research Foundation (Deutsche Forschungsgemeinschaft DFG) via the Emmy Noether Award SCH 1249/2-1. 

\end{document}

%% file: declarations.tex
\def\B#1{\mbox{\boldmath{$#1$}}}

\newcommand{\divg}{{\rm div}}

\newcommand{\nn}{\nonumber}

\newcommand{\mA}{\alpha}
\newcommand{\mB}{{\beta}}

\newcommand{\bu}{\mathbf{u}}

\newcommand{\bw}{\mathbf{w}}

\newcommand{\bv}{\mathbf{v}}

\newcommand{\bx}{\mathbf{x}}

\newcommand{\bg}{\B{g}}
\newcommand{\bJ}{\mathbf{J}}
\newcommand{\bh}{\mathbf{h}}

\newcommand{\bchi}{\boldsymbol{\chi}}

\newcommand{\trho}{\tilde{\rho}}
\newcommand{\bpi}{\boldsymbol{\pi}}
\newcommand{\bPhi}{\boldsymbol{\Phi}}

\def\be{\begin{equation}}
\def\ee{\end{equation}}
\def\ba{\begin{array}}
\def\ea{\end{array}}
\def\bea{\begin{eqnarray}}
\def\eea{\end{eqnarray}}
\def\beas{\begin{eqnarray*}}
\def\eeas{\end{eqnarray*}}
\newcommand{\bseq}{\begin{subequations}}
\newcommand{\eseq}{\end{subequations}}
